\documentclass[prb,aps,floats,floatfix,twocolumn,graphicx,final,nofootinbib]{revtex4}
\usepackage{graphics,graphicx}
\usepackage{amsfonts,amssymb,amsmath}
\usepackage[T1]{fontenc} 

\def\squid{\textsc{squid}}
\def\be{\begin{equation}}
\def\ee{\end{equation}}

\begin{document}

\title{Equilibrium properties of mesoscopic quantum conductors}

\author{L. Saminadayar\footnote{\texttt{QuSpin@grenoble.cnrs.fr}}}
\address{Universit\'{e} Joseph Fourier\\
B.P. 53, 38041 Grenoble Cedex 9, France}
\address{Institut N\'{e}el, Centre National de la Recherche Scientifique\\
25, avenue des Martyrs, B.P. 166 X, 38042 Grenoble Cedex 9, France}

\author{C. B\"auerle\footnote{\texttt{QuSpin@grenoble.cnrs.fr}}}
\address{Institut N\'{e}el, Centre National de la Recherche Scientifique\\
25, avenue des Martyrs, B.P. 166 X, 38042 Grenoble Cedex 9, France}

\author{D. Mailly\footnote{\texttt{dominique.mailly@lpn.cnrs.fr}}}
\address{Laboratoire de Photonique et Nanostructures, Centre
National de la Recherche Scientifique\\
route de Nozay, 91460 Marcoussis, France}

\date{\today}

\maketitle

    


\tableofcontents


\section{Introduction}

The emergence of mesoscopic physics has lead to the discovery of many
striking new phenomena in solid state physics in the last two
decades\cite{imry97}.  This field is also intimately related to the
progress in fabrication techniques: the possibility of creating
objects of sub-micron size has allowed to fabricate and manipulate
conductors which are fully coherent.

In solid state physics, one usually considers \emph{macroscopic}
systems.  This term often refers to the notion of the thermodynamic
limit: the number of particles $N$ and the volume of the system
$\Omega$ both tend to infinity whereas the ratio $n={N/\Omega}$ is
kept constant\cite{ashcroft,kittel}.  This idea is also closely
related to the physical size of the system: the sample is considered
as macroscopic as soon as its size is larger than some characteristic
length, for example the typical distance between two particles,
$n^{-1/3}$.  Below this size, the system is said to be
\emph{microscopic}.

It is well known from our daily experience that macroscopic objects
obey classical mechanics, whereas microscopic ones are governed by
quantum mechanics.  This dichotomy between microscopic and macroscopic
behavior is quite familiar: small particles exhibit wave-like
attributes and they must be described by quantum mechanics that allows
for wave behavior like diffraction or interference.  Electrons have
been observed to interfere in many experiments in vacuum.  However, if
one considers a large number of electrons in a disordered medium, like
a macroscopic piece of metal at room temperature, the conductivity is
described in a classical way \textit{via} the Boltzmann equation which
leads to the Drude formula.

The question is then: is it possible to observe the wave-like
behavior of the electrons in a solid?  Actually, the characteristic
length which is relevant is the length over which the electronic wave
keeps a well defined phase, namely the \emph{phase coherence length}
$l_{\phi}$.  This phenomenon is well known in optics: incoherent light
can not give rise to interference patterns.  

At room temperature, the
phase coherence length of an electron in a metal is of the order of a
nanometer, roughly the $n^{-1/3}$ factor mentioned above.  However, at
low temperature, let us say below $1\,K$, this phase coherence length
increases and may reach several micrometers in metals or even more
than $10$ micrometers in highest quality semiconductor
heterojunctions.  Combined with the progress of fabrication
techniques, this allows to observe the quantum behavior of the
electrons in solids.  It should be stressed that a micron size sample
is in a sense really macroscopic when compared to microscopic scales
(for example the inter-atomic distances): if a mesoscopic sample
behaves like a large molecule to some extent, it still contains a
rather large number of atoms and electrons (more than $10^{20}$). 
However, as the electronic wave function is fully coherent over the
whole sample, this sample is really a \emph{quantum
conductor}\cite{landauer57,landauer70}.

It is important to point out that the physics of such a system is
completely different from the physics of free electrons in vacuum. 
First, as we said, the sample is macroscopic, and some notions,
reminiscent from the standard solid state physics, are still relevant:
the Fermi wave length, Fermi level, Fermi velocity or chemical
potential still make sense; more important, the energy spectrum is
discrete, or, at least, the inter-level spacing $\Delta\approx
L^{-d}$, where $L$ is the size of the system and $d$ the
dimensionality, may become comparable or even smaller than the
temperature $T$.  Secondly, disorder is still present in solids at the
micron scale: thus, the electronic wave propagates in a random medium
and its electronic motion is still diffusive.

It is often tempting to make analogies between interference phenomena 
in mesoscopic physics and optics.  Such a comparison may sometimes be appropriate: for
example, the Young slit experiment and the Sagnac effect are very
similar to some transport experiments on mesoscopic rings like the
Aharonov-Bohm conductance oscillations\cite{sharvin81,webb85,schopfer07} or the
quantization of the conductance which may be understood in the light
of the theory of waveguides\cite{wees88}.  There are, however, two
important differences: first, electrons are fermions, and this
obviously affects strongly the energy spectra of mesoscopic samples 
and consequently their transport and thermodynamic properties; 
secondly, electrons are charged particles and couple to the vector
potential of the electromagnetic field.  This provides a powerful tool
to control interference effects simply by applying a magnetic flux.

This article is organized as follows: in the first part, we will give
the fundamental length scales which are important in mesoscopic
physics.  We will then give an overview of the different materials
commonly used in this field.  In the second part, we will present a
comprehensive overview of the different fabrication techniques. 
Finally, the third part is devoted to the thermodynamics of theses
mesoscopic systems.

\section{Mesoscopic systems}

\subsection{Mesoscopic samples}
\subsubsection{Characteristic lengths}
\paragraph{Mean free path}\label{par.le}

A mesoscopic sample is a disordered sample: even at zero temperature,
electrons are scattered by static defects like impurities, grain
boundaries or the edge of the sample.  Such events are elastic
scattering in the sense that their energy is conserved during the
collision. The disorder just acts like a static, random potential
which adds to the lattice potential.  In such a system, Bloch states
are no more eigenstates, but the system is still hamiltonian.  It
should be pointed out that the translational invariance of
the crystal lattice is destroyed by such defects, but this usually
does not affect the electronic properties of the system.  The typical
length associated with these scattering processes is $l_{e}$, often
called the mean free path.  The time associated with these collisions
is $\tau_{e}$ and they are related \textit{via} the relation
$l_{e}=v_{F}\,\tau_{e}$, $v_{F}$ being the Fermi velocity.

On the contrary, other collisions are inelastic in the sense that the
energy of an electron is not conserved.  Such processes are
irreversible and are related to the coupling of the electrons with
their environment, \textit{i.e.} other electrons, phonons or photons. 
The inelastic length is given by
$l_{in}=\min\{{l_{e-e},l_{e-photon},l_{e-phonon}}\}$.  At high
temperature (typically above $1\, K$), the dominant mechanism is
electron-phonon scattering.  At low temperature, however, the dominant
process is electron-electron scattering\cite{ashcroft,kittel}.

Another important source of decoherence is electron photon 
scattering. 
This is especially the case in micrometer size samples, where small 
dissipation 
(10$^{-15}$ Watts) is sufficient to heat the conduction electrons at 
very low 
temperatures. Extreme care should therefore be taken for external 
radio-frequency 
filtering\cite{zorin95,vion95,glattli97} when working in the 
millikelvin 
temperature range. 

\paragraph{Phase coherence time}\label{par.lphi}

After an inelastic scattering event, the energy of the electron
changes, and the phase of the wave function is randomly distributed
between $0$ and $2\pi$; thus, the quantum coherence is lost and the
phase coherence time is mainly limited by the inelastic time,
$\tau_{\phi}\approx\tau_{in}$. It is important to note, however, that elastic scattering
also leads to dephasing: the wave vector $\vec{k}$ changes to
$\vec{k'}$ after such a diffusion; elastic scattering implies only
that $|\vec{k}|=|\vec{k'}|$, but there is \textit{a priori} no
condition on their respective directions.  The point is that this
dephasing is perfectly deterministic and reproducible: two successive
electrons with the same wave vector $\vec{k}$ will be scattered and dephased 
in exactly the same way, which could be calculated, if the scattering potential were
known.  The phase coherence is thus preserved, and interference
effects are not destroyed.  On the contrary, inelastic scattering
depends on the state of the environment the electron interacts with at
the time of the interaction. In this case, the dephasing is random and
the phase coherence lost.  This is why, at room temperature, the
dominant scattering process is the electron-phonon scattering, and the
phase coherence length is very short, typically
$l_{\phi}\approx1-10\,nm$. In the frawework of Fermi liquid theory, 
the available phase space at low tempertaure tends to zero. As a 
consequence, electron-electron, electron-photon and
electron-phonon couplings all tend to zero and hence, the phase coherence length
should diverge\cite{altshuler}.  Recent experiments, however, seem to
show that this is not the case.  Presently, there is still an ongoing
debate concerning this point, and we will not address this issue in
this article\cite{mohanty_prl_97}.

Finally, it is important to mention magnetic impurities.  As they are
static defects at low temperature, scattering by magnetic impurities
is elastic as the energy of the electron is conserved.  However, the
electronic spin is flipped in such a collision, and the phase
coherence may be lost.  The exact effect of magnetic scattering on the
phase coherence time, especially when entering the Kondo regime, is
far from being understood\cite{schopfer03,bauerle,mallet}.  In this article, we
will not elaborate on this point and consider only systems containing 
no
magnetic impurities.

\paragraph{Thermal diffusion length and the Thouless
energy}\label{par.lT}

At distances beyond the elastic mean free path
$l_{e}$, electrons propagate in a random medium.  This diffusive
nature of the movement is characterized by the diffusion coefficient
$D={1\over d}{v_{F}}^2\tau_{e}={1\over d}v_{F}l_{e}$ where $d$ is the
dimensionality of the sample.  To propagate over a distance $L$, an
electron then needs a diffusion time $\tau_{d}=L^2/D$.  In a
semi-classical picture, each diffusion path $l$ is characterized by a
probability $\Psi_{l}=|\Psi_{l}|\exp(iS_{l}/\hbar)$, where
$S_{l}=\int_{l}\vec{k}d\vec{l}-Et_{l}$ with $E$ being the energy of
the electron and $t_{l}$ the diffusion time along the path $l$.  Over
the whole sample of size $L$, this diffusion time is then simply
$\tau_{D}$.  If one considers an energy range larger than
$2\pi\hbar/\tau_{D}$, the phase of the electrons in this energy range
will be distributed between $0$ and $2\pi$, and interference effects
will not be observable anymore.  This defines the Thouless energy (or
correlation energy) $E_{c}=h/\tau_{D}=hD/L^2$.  When the energy range
involved is larger than $E_{c}$ (\textit{e.  g.} when $k_{B}T\geq
E_{C}$), interference effects do \emph{not} disappear, they are simply
no more observable\cite{thouless72}.  If the size of the sample is
smaller than $l_{e}$, the time for an electron to travel across the
sample becomes simply $L/v_{F}$, and the Thouless energy simply
expresses as $E_{c}=hv_{F}/L$.

\subsubsection{Disorder configurations}
In a macroscopic sample, one usually characterizes the disorder by some
characteristic length, say the elastic mean free path $l_{e}$.  Such a
parameter is relevant when considering the disorder from a ``global''
point of view.  From a mesoscopic point of view, things may
be quite different: the electronic wave functions are fully coherent
over the whole sample, and the acquired phase depends on the precise
path one electron follows.  Thus the interference pattern depends on
the \emph{microscopic} disorder configuration of the sample.  Moving
even a single impurity affects drastically the electronic properties
of the sample.  That is why two samples identical from a macroscopic
point of view may behave in a completely different way, due to
their microscopic individuality (their \emph{fingerprints}).  This
phenomenon is equivalent to the speckles observed when a coherent
light beam diffracts in a random medium.

\subsubsection{Quantum coherence and the effect of AB flux}
The most important parameter that physicists can use to probe a
mesoscopic sample is the magnetic flux.  As an electron is a charged
particle, it couples to the vector potential $\vec{A}$ (the momentum
changes as $\vec{p}\rightarrow\vec{p}+e\vec{A}$ in the hamiltonian,
with $e$ the charge of the electron) \emph{even if the magnetic field
$\vec{B}$ is zero}
($\vec{B}=\vec{\bigtriangledown}\times\vec{A}=\vec{0}$).  Note however
that as the field is zero, or at least very weak in all the
experiments\footnote{Except in the case of the Quantum Hall Effect,
that we will not address in this article.}, the effect of the magnetic
field on the trajectories of the electrons is negligible.  When
propagating along a path $i$, the wave function $\Psi$ acquires a
phase simply given by
$S=\int_{i}(\vec{k}(\vec{r})+e\vec{A}(\vec{r}))d\vec{r}$.  The first
term is simply the equivalent of the optical path, whereas the second
one characterizes the quantum coupling of the charge with the magnetic
flux.  This shows how applying a small magnetic field can indeed
control the interference pattern of a mesoscopic
sample\cite{chakravarty86}.  There is no equivalent of such a
possibility in optics: this is a powerful way to play with the
quantum, wave-like nature of the electrons.

\subsection{Materials}
\subsubsection{Ballistic vs diffusive vs localized}
The different length scales for a mesoscopic sample are the Fermi
wavelength $\lambda_{F}$, the elastic mean free path $l_{e}$ and the
size of the sample $L$.  The ratio between $\lambda_{F}$ and $l_{e}$
characterizes the strength of the disorder: for $\lambda_{F}\ll l_{e}$
(or, equivalently, $k_{F}l_{e}\gg 1$ or $h/\tau_{e}\ll E_{F}$), the
disorder is said to be ``weak'', whereas for $k_{F}l_{e}\ll 1$ the
disorder is said to be ``strong''.

Considering the ratio between these different length scales, one
can distinguish different regimes for a mesoscopic sample.

\paragraph{Ballistic regime}\label{par.bal}
($\lambda_{F}\ll l_{e}$ and $L\leq l_{e}$)

In this regime, the
disorder is very weak and the elastic mean free path is of the order
of the size of the sample.  In this case, the phase coherence length
is mainly limited by electron-electron collisions.  The trajectories
of the electrons is mainly governed by the shape of the sample, implying
that the reflections at the edges of the samples are specular.  In 
this
case, transport properties as well as equilibrium properties depend on
the shape of the sample.  Such systems are powerful tools to probe
the energy spectra of quantum billiards.

\paragraph{Diffusive regime}\label{par.dif}
($\lambda_{F}\ll l_{e}\ll L$)

In such systems,
electrons experience a large number of collisions during the traversal
of the sample.  Their movement is rather a brownian motion, a random
walk between impurities.  The phase coherence length is then given by
$l_{\phi}=\sqrt{D\tau_{\phi}}$.  In this regime, the exact shape of
the sample does not affect its electronic properties; only its size is
relevant.

\paragraph{Localized regime}\label{par.loc}

In the case of a strong disorder, Anderson has suggested that each
electron is confined in a part of the sample and can not travel
through it: its wave function is exponentially decreasing on a length
scale $\xi$ and the electron is localized in a domain of size $\xi^d$
with $d$ the dimensionality of the sample, and the sample becomes an
insulator\cite{anderson58}.  For $d=3$, there is a critical value for
the disorder below which the sample becomes insulating and one
observes a metal to insulator transition.  For $d=1$ and
$d=2$ on the other hand, electrons are localized for an arbitrary small
disorder\cite{abrahams}.  Recent experiments, however, show that there is
indeed a metal to insulator in some two dimensional electron gas.  As
both the experimental and theoretical situation are at least unclear,
we will not address this topic in this article\cite{pudalov}.

Two limits are then to be considered\footnote{Note that one always has
$\xi\geq l_{e}$.}: when $\xi\prec L$, electrons are confined in some
regions of the sample, and conduction occurs by hopping from domain to
domain.  On the other hand, when $\xi\succ L$, localization domains are
larger than the sample: electrons are indeed localized but they can
still explore the whole sample.

\subsubsection{Dimensionality}
One defines the dimensionality of a sample by comparing its size with
the intrinsic characteristic lengths\cite{beenaker91}.  Usually, the
most relevant length scale is the Fermi wavelength.  Considering a
rectangular sample of sizes $L_{x}$, $L_{y}$ and $L_{z}$, with
$L_{x}\prec L_{y}\prec L_{z}$, one has:
\begin{eqnarray*}
    \lambda_{F}\ll L_{x} \prec L_{y} \prec L_{z} & : & \textrm{3
    D  (bulk samples)} \\
    L_{x} \leq \lambda_{F}\ll L_{y} \prec L_{z} & : & \textrm{2
    D (films) } \\
    L_{x} \prec L_{y} \leq \lambda_{F}\ll L_{z} & : & \textrm{1
    D (quantum wires)} \\
    L_{x} \prec L_{y} \prec L_{z} \leq \lambda_{F} & : & \textrm{0
    D (quantum dots)}
 \end{eqnarray*}
 
 Such a definition is certainly the most relevant from a microscopic
 point of view.  Note however that when considering transport
 properties, and due to the quantum nature of a mesoscopic conductor,
 one can also define the dimensionality of a sample by comparison with
 the phase coherence length:
 \begin{eqnarray*}
l_{\phi} \ll L_{x} \prec L_{y} \prec L_{z} & : & \textrm{$3D$} \\
L_{x} \leq l_{\phi} \ll L_{y} \prec L_{z} & : & \textrm{$2D$} \\
L_{x} \prec L_{y} \leq l_{\phi} \ll L_{z} & : & \textrm{$1D$}
 \end{eqnarray*}
 
\subsubsection{Metals}
Metals have a high charge carrier density of about $10^{22}\,cm^{-3}$. 
Because of this high carrier density, the Fermi wavelength is very
short, in the range of the Angstr\"{o}m.  Moreover, it is impossible
to use gates to modulate this electron density (too important voltage
would be necessary in the case of metals).  Another consequence is
that the Coulomb interaction is very efficiently screened on the scale
of the Thomas-Fermi vector, $q_{TF}=2\pi e^2/\rho_{0}$, with
$\rho_{0}$ the carrier density at the Fermi level.  Even if metals can
be very pure from a chemical point of view, the intrinsic disorder
usually makes them diffusive conductors.  The elastic mean free path
$l_{e}$ is of the order of $1-100\, nm$, and the phase coherence
length $l_{\phi}$ in the order of the micrometer\footnote{In extremely
clean metals, obtained by Molecular Beam Epitaxy (MBE), the phase
coherence length can reach $\approx 20\mu m$ at best.}.

At low temperature, some metals become superconductors.  This provides
a new degree of freedom, and a wide variety of mesoscopic effects.  In
particular, the superconducting state is quite different on a
mesoscopic scale as compared to its macroscopic equivalent.

\subsubsection{Semiconductors}
\paragraph{Bulk semiconductors}\label{par.bul}

In semiconductors, the carrier densities can range practically between
$10^{14}\, cm^{-3}$ and $10^{19}\, cm^{-3}$.  Moreover, this density
can be controlled using metallic gates deposited at the surface of the
sample, or simply by varying the doping concentration.

In the case of very pure semiconductors, \textit{eg} those obtained by
Molecular Beam Epitaxy (MBE), the elastic mean free path is basically
limited by the distance between two doping impurities.  This leads
easily to $l_{e}$ of $\approx 100\, nm$, whereas $l_{\phi}$ is of the
order of several micrometers. Finally, another important difference
between metals and semiconductors is that in the latter, the effective
mass of the electrons, which is related to the band structure, can be
very small.

\paragraph{Heterojunctions}\label{par.het}

To reduce the dimensionality of a conductor, one may reduce the
thickness of the film itself.  However, it is quite di?ficult to
obtain real two dimensional conductors on the scale of the
Fermi wavelength $\lambda_{F}$.  An alternative way consists in
playing with the band structure of two different semiconductors. 
Using the impressive control of growth offered by the Molecular Beam
Epitaxy (MBE), it is possible to grow two different semiconductors on
top of each other, especially if their lattice parameters are
matched\cite{esaki69}.  The most common example is GaAs and GaAlAs
(III-V heterostructures), but there also exist II-VI heterostructures
(CdTe/HgCdTe) or even IV-IV heterostructures (SiGe).

The different band structure, mainly the energy gap and the work
function\footnote{For example in GaAlAs\cite{asga1,asga2}, the gap
$E_{g}$ varies linearly with the concentration of aluminium as
$E_{g}=1,424+1,247x$ ? $300\, K$, $x$ being the concentration in
aluminium.  Moreover, the mismatch in the lattice parameter does not
exceed $0,3\%$.}, causes changes in the charge transfer between the
two adjacent materials in order to equalize the electrochemical
potentials.  Electrons are attracted to the remaining holes and the
dipole layer formed at the interface leads to the band bending at the
vicinity of the interface.  True two-dimensional electron (or hole)
gas at the scale of $\lambda_{F}$ can be formed using this
technique\cite{ando82}.

The spatial separation between charge carriers and doping impurities
leads to very high mobility materials\footnote{The highest mobility
achieved in GaAs/GaAlAs heterostructures\cite{heiblum97} is $14\cdot
10^6\,cm^2\,V^{-1}\,s^{-1}$.  Mobilities of $\approx
10^6\,cm^2\,V^{-1}\,s^{-1}$ are currently achieved in this 
material.}. 
The electronic density is typically in the range of $10^{11}\,
cm^{-2}$, leading to a relatively large Fermi wavelength, of the 
order of 300 $\AA$.  This large Fermi wavelength allows to create
easily true $1D$ or $0D$ structures.  Moreover, the use of
electrostatic gates on the top of the sample allows to deplete the 2D
electron gas underneath.  Using this technique, one can modulate
\textit{in situ} and in a reversible way the shape of the $2D$
electron gas, allowing to create a wide variety of quantum devices,
like quantum wires or quantum dots\cite{beenaker91}.  Moreover, the
edges defined by electrostatic gates are by far less rough than those
produced by etching techniques.

\section{Sample fabrication techniques}

\subsection{The size to reach}

Typically one wants to be able to taylor samples with a size smaller
than one micrometer.  But the smallest size is not the ultimate goal since
the roughness of the edge may play an important role.  In metals, the
Fermi wavelength $\lambda_F$ is very short and the roughness is always
much larger than the Fermi wavelength.  But in semiconductor samples,
where $\lambda_F$ can be several tens of nanometer, the roughness can
be of the same order.  In this case one wants the edges to be defined
with a precision much smaller than $\lambda_F$.  As discussed in the
previous paragraph, the dimensionality of the sample depends strongly
on the physics involved.  For interference effects, the phase
coherence length is the characteristic length which is of the order of
1nm at low temperature for a good metal and more than 10$\mu$m for
high quality 2DEG. So typically one wants to be able to fabricate
samples with a width smaller than 100nm for metal structures and a few
hundred nanometers for semiconducting ones.

\subsection{Nanofabrication technique}

\begin{figure}[tbp]
\centerline{\includegraphics*[width=8cm]{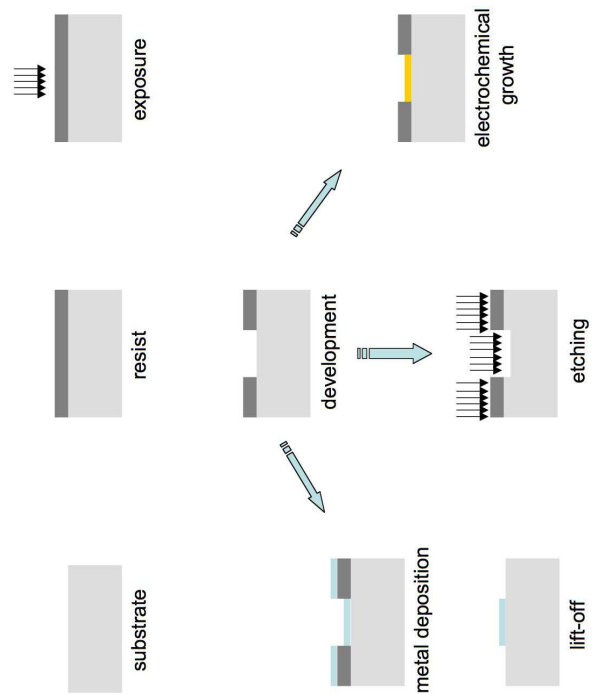}}
\caption{\textsc{Sem} A typical flowchart for a fabrication process.}
\label{fabrication}
\end{figure}

First of all, let us recall a standard process flow. 
Figure~\ref{fabrication} resumes the main steps one must follow.  The
starting material is the substrate, that can be the system one wants
to pattern or just a flat and neutral surface used as a support.  By
spinning, the substrate is coated with a layer of resist.

The resist is a material sensitive to irradiation.  After exposure,
the resist is developed and the exposed (non exposed) area will be
cleared off for the case of positive (negative) resist.  A rich
variety of processes can be done after the lithography.  A commonly
used process is the lift-off technique.  In this case one covers the
whole patterned substrate with a metal for instance.  The resist is
then completely removed by rincing it with a strong solvant.  Only the
part which had been previously patterned will be covered by the metal,
so to say one has replaced the design on the resist by a solid pattern
made of metal.  This metal can be just the structure wanted or can be
used as a mask for a subsequent etching process.  Other processes can
also be used such as ion implantation, electrochemical growth etc.

\subsection{Optical lithography}

Optical lithography is the dominant lithography in industry.  With
this technique UV light is shed through a mask, which contains the
drawing information, on a resist.  The resolution is mostly limited by
the diffraction and hence depends on the wavelength of the light. 
This explains why short wavelength are employed.  Optical lithography
which started with UV (400nm to 366nm) is now in the DUV range (248nm
to 193nm) and EUV (13nm) is the next predicted range.  DUV lithography
can reach the sub 100nm range but with a complexity and costs which
are too high for any scientific laboratory.  For instance, the
complexity of masks which use phase shift techniques to overcome
diffraction make them difficult and hence very expensive to produce. 
Only mass production can afford such high costs.  Refractive optics
are presently not available in the EUV range.  The fabrication of
reflective optics at this wavelength is also very delicate and it is
hopeless that this technique will be unexpensive for laboratory use. 
Classical optical lithography, on the other hand, which use quartz
plate mask directly pressed onto the resist with a standard DUV light
is not able to produce sample with a sub 100nm resolution.

\subsection{Electron beam lithography}

The possibility to finely focus an electron beam has been exploited in
electron microscopy since long time ago.  Starting in the sixties,
focused electron beam has been used to expose a resist and a 0.1$\mu$m
resolution was readily obtained.  Ten years later, a 10nm wide line
was demonstrated using an inorganic resist.  Unfortunately this
technique is essentially sequential: the electron beam is scanned
pixel by pixel on the resist to draw the entire design, hence the
process is too slow to be included in an industrial processes.  It is
on the other hand the perfect technique for the laboratory.  An
advantage of this maskless technique is its versatility.  The drawing
can be easily changed on a computer with no additional cost.

In the following we detail the electron beam lithography to explain
the resolution and limitations of this technique.

\subsubsection{Resolution and proximity effect} 

Most of the resists employed in nanotechnology are polymers.  The
effect of the electron is to break the chain, hence leaving a polymer
with a small chain giving a better solubility.  This resist is then
sensitive to a very small energy compared to the one of the electron
beam.  Typically one needs 10 eV to break a polymer chain whereas for
technical reason the focused electron beam is accelerated at several
tens of kilovolts.  It is then important to know how the electrons
diffuse into the resist and loose their energy in order to understand how the 
resist is affected.

\begin{figure}[tbp]
\centerline{\includegraphics*[width=8cm]{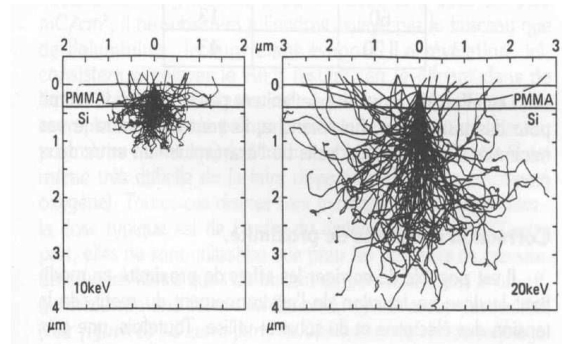}}
\caption{\textsc{Sem} Monte Carlo simulation of electron trajectories 
in Silicon substrate covered with PMMA at 10 keV and 20 keV. After 
ref.\cite{Kyser}.}
\label{insol}
\end{figure}

An analytical treatment is quite complex, especially in three
dimensions.  Monte Carlo simulations are widely used to follow the
electron trajectory.  Figure \ref{insol} shows electron trajectories
obtained for electrons with energies of 10kV and 20kV in a silicon
substrate covered with 400nm of PMMA resist.  The effects of electron
diffusion is twofold: first, a forward diffusion which enlarges the
spot in the resist is observed. Secondly, a backscattering diffusion, mainly from
electron diffused in the substrate back into the resist but far from
the initial impact of the electron, takes place.  This latter effect, known as
proximity effect, has important consequences as we will see later.  The
energy of the beam is quite important as can be seen in figure
\ref{insol}: higher energy decreases the forward scattering angle and
shrink the effective beam spot.  On the other hand, the electron
penetrates more deeply into the substrate as their energy increases and
are backscattered at larger distance from the impact.  In other words,
higher energy dilutes the proximity effect.  This is the reason why
recent electron beam machines use a 100kV source.

The total dose received by the resist at one point depends on the
exposed dose at that point, but also on the vicinity around this point. 
Hence a large square uniformly exposed, for instance, will be more dosed
in the center than on the edge.  It is also very difficult to exposed
two large patterns close to each other. The gap between these two 
patterns being exposed by proximity, may result in an unwanted 
connection between them.  Arrays of lines with a very small pitch are
also very difficult.  It is possible, however, to correct the dose at
each point by calculating the proximity effect of the overall pattern. 
Softwares have been developed for that purpose, but cannot completely
cure the effect of the diffusion, since it may require a negative dose at
certain points!

The problem of proximity effect arises from the
sensitivity of organic resist to small energies.  It is thus natural to
try to use resist which needs higher energy to be exposed.  This is
the case of inorganic materials \textit{e. g.} NaCl, AgF$_2$, Al$_2$O$_3$,... 
Such inorganic resist have been used to demonstrate the finest lines
obtained by e-beam lithography, around 1nm.  The beam energy in that
case gives rise to the partial or total sublimation of the resist. 
For instance on AlF$_3$, the electron energy evaporates Fluor leaving
a layer of Aluminium.  Hydrocarbon films have also been used where
under irradiation a polymerisation takes place.  In most cases,
the dose necessary to expose these type of resists is orders of
magnitude higher than with conventional resist.  The total time to
expose the pattern can reach non reasonable values.  Furthermore, this
type of resist can be used only with thin layers which enable any
lift-off process.  Another possibility to avoid proximity effect is to
use a very small energy.  In this case, however, it is very difficult
to focus the beam in conventional electron optics due to chromatic
aberration.  Another drawback is the forward scattering which rapidly
enlarges the beam in the resist.

\begin{figure}[tbp]
\centerline{\includegraphics*[width=8cm]{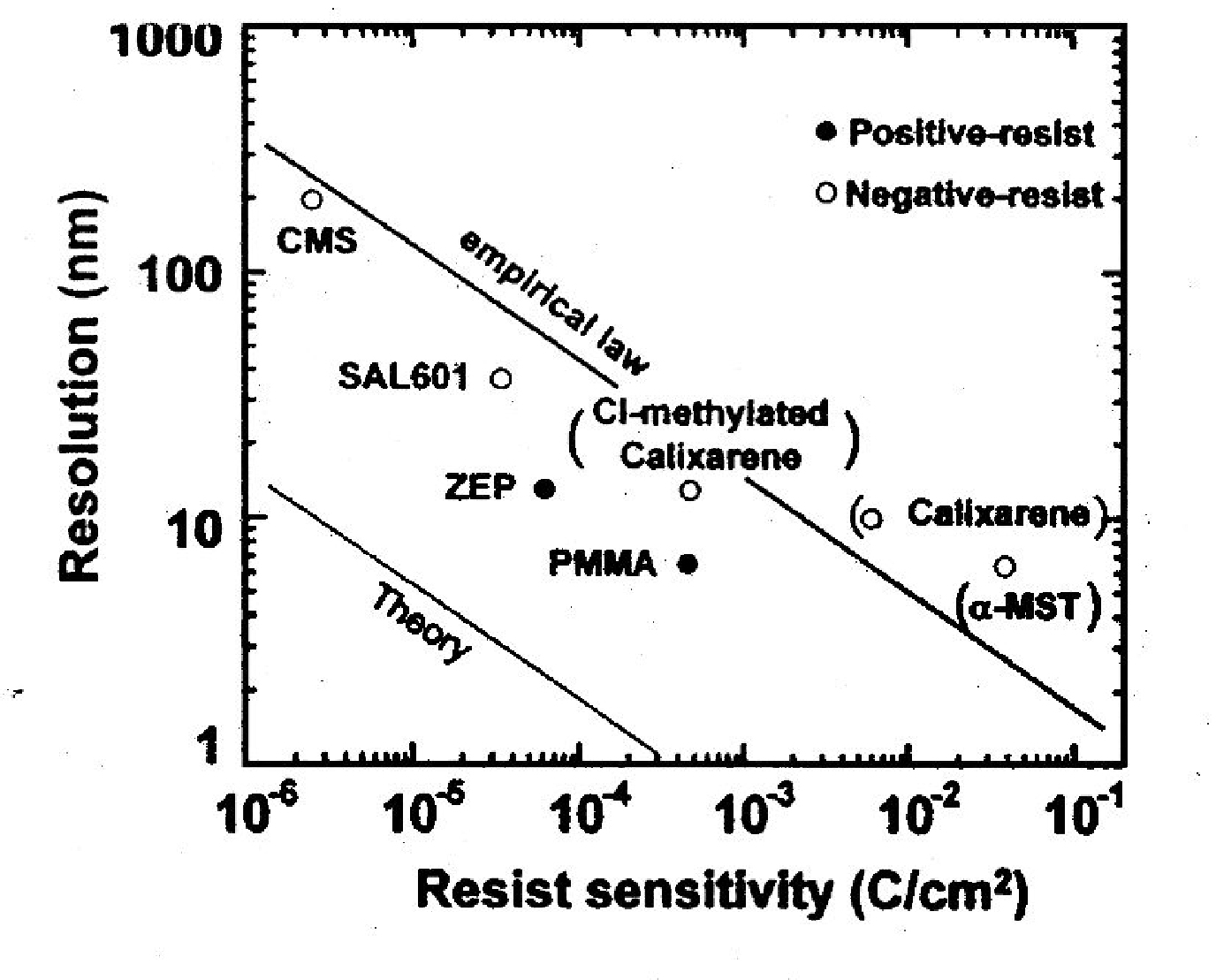}}
\caption{\textsc{Sem} Resolution and sensitivity of various organic
resists positive and negative for electron beam lithography. 
PolyMethylMetAcrylate (PMMA) is presently the best organic resist and
is most widely used in scientific laboratories around the world.  An
example of sub 10 nm lines of PMMA with a 50 nm pitch is shown in
figure \ref{pmma}.}
\label{resist}
\end{figure}

\begin{figure}[tbp]
\centerline{\includegraphics*[width=8cm]{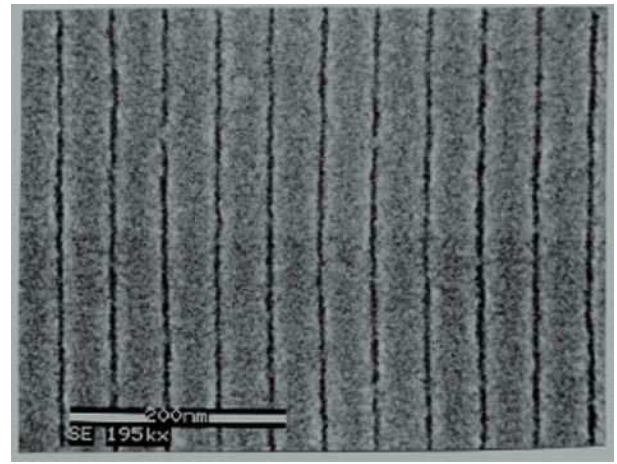}}
\caption{\textsc{Sem} The image shows an example of sub 10nm lines
with a 50nm pitch\cite{vieu2}.}
\label{pmma}
\end{figure}

Figure \ref{pmma} shows the best resolution obtained with different
organic resists and the dose needed with e-beam lithography.  One
should keep in mind that the maximum current available in an e-beam
system with a small spot size (less than 10nm) is about 100pA with a
field effect gun source.  This means that exposing an area of
100$\mu$m$\times$100$\mu$m at a nominal dose of 10$^{-1}$C/s takes
more than 27 hours!

An electron microscope with a computer assisted deflection system is
the basic tool for e-beam lithography.  It is enough to make simple
patterns in a single field.  The available field size depends on the
desired resolution.  Lens aberrations induce severe distorsions at the
edge of the field which depend on the field size.  With a conventional
microscope, a 50$\mu$m$\times$50$\mu$m is usually the maximum size one
can afford to produce sub 100nm structures.  The nanostructure has
then to stand within a single field since there is no possibility to
displace the sample holder with enough accuracy to stitch with the
previous writing field.  The stability of the electron column is also
a problem for long time exposure.  Dedicated machines have been built
to overcome the difficulties mentioned above with conventional
electron microscopes.  They include a laser interferometry controlled
stage with an accuracy better than 1nm to measure mechanical
displacement.  A feedback to the electron deflection is usually chosen
for the field alignment.  The overall field stitching accuracy is of
the order of 20nm.  Using patterned marks on the sample it is also
possible to align several layers of lithography.  The mark detection
system combined with the laser interferometry allows also to calibrate
the deflection amplifier and to correct field distorsion.  It is
simply done by moving a mark at different positions in the field.  The
exact position of the mark is known using the laser interferometry and
is compared to the position of the mark obtained by deflecting the
beam.  All these essential features explain that there is at least one
order of magnitude in the price of such a machine compared to a
standard electron microscope.

\subsection{Other charged particle lithography}

Focused ion beam lithography arose rapidly after electron beam
lithography as a good candidate for nanofabrication.  Ions offer
several advantages compared to electrons: first they deliver very
quickly their energy and consequently a much smaller dose is
necessary.  Secondly the throughput is much better and the proximity
effect is much smaller.  In addition, ions can directly erode the
material and is a resist free process which can be very interesting
for materials which are sensitive to pollution by organic materials.

\begin{figure}[tbp]
\centerline{\includegraphics*[width=8cm]{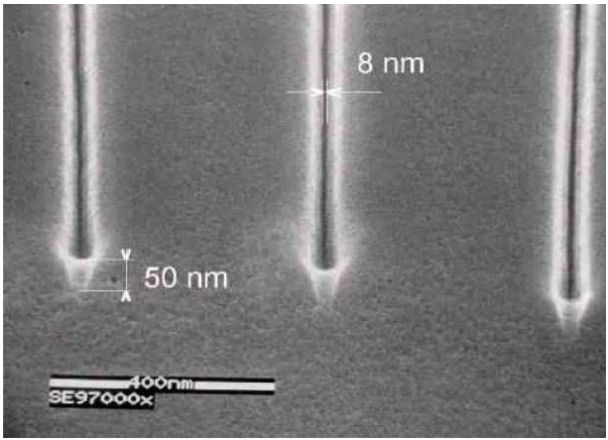}}
\caption{\textsc{Sem} 8nm line in GaAs produced using a 20kV Gallium focused 
ion beam. After ref.\cite{Gierak}.}
\label{GaAs_line}
\end{figure}

In this etching mode by varying the dose, it is also possible to produce
three dimensional structures: the paradigm of nanofabrication. 
Finally, at higher energy one can locally implant atoms.  On the other
hand, ion lithography did not take an important place among the
nanofabrication techniques.  The major reason is the difficulty to
produce fine spots with enough current and a good stability.  Recent
progress in ion optics and ion source technology succeed in producing
sub 10nm spot sizes with a particle density sufficient for etching. 
Figure \ref{GaAs_line} shows an 8nm line produced by such high
resolution ion system\cite{Gierak}.

\subsection{Near field techniques}

Shortly after their discovery, near field technique have been used to
produced nanostructures.  The ultimate resolution has been obtained by
the IBM group who wrote the acronym of their company using Xenon atoms
with a scanning tunneling microscope (STM)\cite{ziegler}.  But such
nanostructures are unfortunately very volatile.  STM can also be used
in a more conventional way as a source of focused electrons.  Indeed
the size of the electron spot of a STM is approximately on the order
of the sample to tip distance and a 10nm spot size can readily be
obtained.

Exposure of resist is complicated by the fact that most of the resist
is non conducting.  Another problem due to the proximity of the tip
to the resist is the swelling of the resist under irradiation that can
damage the tip.  STM lithography is more used without resist by
electrochemical process.  For instance, it is possible to remove a
group of atoms by applying a pulse on a gold surface\cite{zzwang}. 
Local oxidation is also an electrochemical process that is widely used
with atomic force microscopy (AFM) lithography.  The native water film
on the surface of a sample at room temperature is the medium for this
anodization process.  In GaAlAs/GaAs samples, it is possible for
instance to locally oxidize the surface and the oxide formation
destroys the two dimensional electron gas beneath.  Several mesoscopic
structures have been produced with this technique\cite{ensslin}. 
Another example is the use of Niobium which can be anodized\cite{bouchiat}.  One of the advantage of this technique is that,
using a small voltage on the tip, one can visualize the structures
obtained at higher voltage.  Usually one is limited to a small writing
field because of the hysteresis of the piezo-displacement.  In most
cases this technique is combined with other techniques like optical
lithography. The possibility of visualization, with the AFM in the
non writing-mode allows for the alignment of the two steps.


\section{Persistent currents: theoretical aspects}
Usually, one considers \emph{transport} properties of quantum
conductors, measured by connecting voltage and current
probes to the sample.  In this case, however, two important 
properties of such a measurement must be pointed out:
\begin{itemize}
    \item First, the strong coupling between these voltage and current
    probes certainly affects the quantum properties of the sample and
    thus the measurement itself.  \item Secondly, in a transport
    experiment, one only probes an energy range $eV$ around the
    Fermi energy, with $e$ being the electron charge and $V$ the 
applied voltage.  That is why one can not access the entire energy 
spectrum
\end{itemize}
It is therefore very interesting to deal with the
\emph{equilibrium} properties of mesoscopic samples.  It has to
be stressed, that such experiments in the field of mesoscopic physics 
are by far much more difficult than transport experiments.  This is 
why there is only a very small number of experimental data available. 

The existence of persistent currents has been first suggested by
London in 1937\cite{london37}, in his studies on the diamagnetism of
aromatic rings (benzene rings).  In 1938, Hund suggested that such an
effect could be present in clean, metallic samples at low
temperature\cite{hund38}.  The amplitude of the persistent currents 
has been first calculated by Bloch and Kulik in the case of a clean,
$1D$ ring\cite{bloch65,kulik70}, but their existence in a real,
diffusive $3D$ metallic ring has only been predicted by B\"{u}ttiker,
Imry and Landauer\cite{buttiker83} in 1983.

It is important to note that the persistent current we are 
considering here is a
\emph{non-dissipative} current flowing in a \emph{non-superconductor} 
ring.  
Another interesting point is that
persistent currents and orbital magnetism are two phenomena completely
equivalent from a physical point of view.  Only the geometry of the
sample makes one term or the other more ``intuitive''.

\subsection{A simple picture: the $1D$ ballistic
ring}\label{simple-model} The simplest model for the persistent
currents is the case of a pure, $1D$ metallic ring, without disorder. 
Although somewhat ``academic'', this example allows to present the 
main idea of the problem.  Let us consider a ring of perimeter $L$ 
pierced by
a magnetic flux $\Phi$. We take the ring to be smaller than the 
phase coherence length  $l_{\phi}$ and we neglect its self-induction.
The hamiltonian for the electrons of the rings is then simply given 
by: 
 \begin{equation}
{\mathcal{H}}=\frac{1}{2m}\left[\overrightarrow{p}-e\overrightarrow{A}\right]^2+V(\overrightarrow{r})
\label{eq:hamiltonien}
\end{equation}
where $\overrightarrow{p}$ is the momentum of the electron, $e$
its charge, $\overrightarrow{A}$ the vector potential, and
$V(\overrightarrow{r})$ the periodic potential of the lattice.  A 
simple gauge transformation
$\displaystyle\overrightarrow{A}\Rightarrow \overrightarrow{A} +
\overrightarrow{\nabla}
\left(\int\overrightarrow{A}.\overrightarrow{dl}\right)$, leads to the
hamiltonian of free electrons ${\mathcal{H}_{0}} = p^2/2m + V$,
whereas a phase is added to the wave function $\Psi$:
$\displaystyle\Psi(x) \Rightarrow \Psi(x)\exp\left(ie/\hbar
\int\overrightarrow{A}.\overrightarrow{dl}\right)$.  This wave
function
then obeys the new boundary conditions\cite{byers61}:
\begin{equation}
\displaystyle \Psi (x+L) = \Psi (x) \exp\left(i \frac{e}{\hbar}
\oint \overrightarrow{A}.\overrightarrow{dl}\right) = \Psi (x) \exp\left(2i \pi \frac{\Phi}{\Phi_{0}}\right)
\label{eq:condlim}
\end{equation}
where $\Phi_{0}=h/e$ is the flux quantum.  These boundary conditions
also lead to a new quantization for the wave vector:
$k=2\pi/L(n+\Phi/\Phi_{0})$.

It should be noted that in this case, the boundary conditions
can be controlled simply by varying the magnetic flux.  Moreover, such
boundary conditions show that wave functions, eigenenergies as well as
\emph{any thermodynamic property of the system are periodic with
magnetic flux\cite{cheung88}, with periodicity $\Phi_{0}=h/e$}.
\begin{figure}[tbp]
\centerline{\includegraphics*[width=8cm]{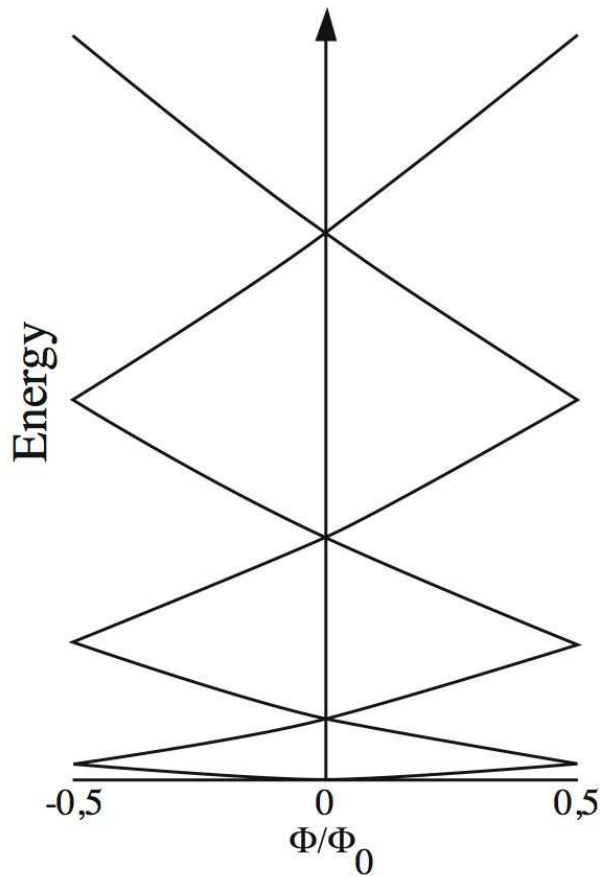}}
\caption{Energy spectrum of a pure, ballistic one-dimensional ring as 
a function of the magnetic flux.}
\label{Spectre1D}
\end{figure}

In analogy with known results on Bloch's states, one can define a
velocity for each energy level\cite{buttiker83,byers61,cheung88,landauer85b}:
\begin{equation}
    v_{n} =\frac{1}{\hbar}\frac{\partial\varepsilon_{n}}{\partial k} =
    \frac{L}{e} \frac{\partial\varepsilon_{n}}{\partial \phi}
    \label{eq:v}
\end{equation}
This velocity is equivalent to a current which is given by:
\begin{equation}
    i_{n}=-\frac{ev_{n}}{L}=-\frac{\partial \varepsilon_{n}}{\partial
    \phi}
    \label{eq:i}
\end{equation}

At zero temperature, the net current is then simply the sum of the 
currents
carried by the $N$ levels:
\begin{equation}
    I_{N}=\sum_{n=0}^{N}i_{n} = \sum_{n=0}^{N}-\frac{\partial
    \varepsilon_{n}}{\partial \phi}=\frac{\partial E(N,\Phi)}{\partial
    \phi}
    \label{eq:In}
\end{equation}
where $E$ is the total energy of the $N$ electrons of the ring. 
However, as can be seen on figure \ref{Spectre1D}, two consecutive
levels carry two currents of the same amplitude, but of opposite sign:
the net current is then simply given by the last occupied
level\cite{cheung89}, \textit{i. e.} the Fermi level.  We thus obtain 
for the amplitude of the persistent current:
\begin{equation}
    \displaystyle I_{0}=\frac{ev_{F}}{L}
    \label{eq:I0}
\end{equation}
In this expression, $v_{F}/L$ is simply the time needed for an
electron to perform one turn around the ring.  It should be noted that
this expression can be rewritten as a function of the Thouless energy
$hv_{F}/L$:
\begin{equation}
I_{0}=\frac{ev_{F}}{L}=\frac{hv_{F}}{L}\frac{e}{h}=\frac{E_{c}}{\phi_{0}}
\label{eq:I02}
\end{equation}

It should be stressed that the persistent current depends strongly on
the number of electrons in the ring and on its parity, both in
amplitude as well as in sign: for $N$ even, the current is
paramagnetic, whereas for $N$ odd, it is diamagnetic\footnote{This is
the case for aromatic rings: for benzene ($N=3$) for example, the 
persistent current is diamagnetic.}.

This very simple approach for the pure $1D$ ring allows to give a 
good estimate for the order of magnitude of the persistent current.  
Moreover, the
main features, such as the dependence on the parity of the number of
electrons remain true, even in the more realistic $3D$, disordered
ring.

\subsection{Realistic ring}
\subsubsection{Introduction}
In this section, we will consider the case of diffusive, $3D$ rings. 
As we have stressed above, each sample is unique due to its
specific disorder configuration.  To take into account this unicity, 
we will consider a large number of rings, which is equivalent to averaging
over disorder configurations: one obtains the \emph{average 
current}.  As we will see, this average current is measured in many rings
experiments.  Fluctuations from this average value are also of
interest, as they are accessible experimentally: this is called the
\emph{typical current}.  This typical current is a good approximation
of the current measured in single ring experiments.

\begin{figure}[tbp]
\centerline{\includegraphics*[width=8cm]{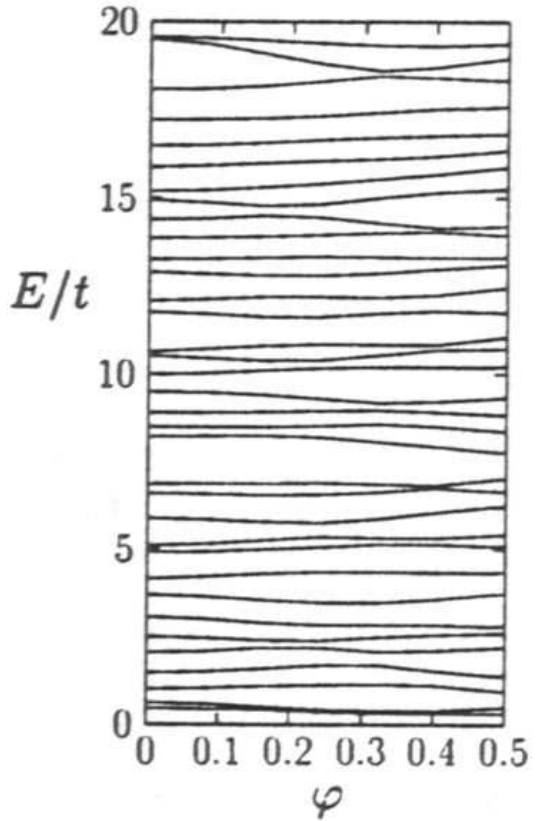}}
\caption{Energy spectrum of a real, three dimensional diffusive ring 
as a function of the magnetic flux (after 
ref.\cite{montambauxCours}.)}
\label{Spectre3D}
\end{figure}

Another important property of the $3D$ case is the \emph{spectral
rigidity}.  In $1D$, we have seen that two successive energy levels
have opposite slopes.  In $3D$, two successive levels repel each
other: this correlation between energy levels leads to a correlation
in the slopes of the energy levels\cite{brody81}, as can be seen in
figure \ref{Spectre3D}.  The slopes of two successive levels are
almost identical, and this correlation extends over an energy range
corresponding to the Thouless (or correlation) energy $E_{c}$.  In
other words, it is necessary to explore an energy range $E_{c}$ to
find a level of opposite slope\cite{bouchiat91,gefen92}.  An important
consequence of that is that the sign of the persistent current still
depends on the number of electrons, but one has to add $E_{c}/\Delta$
electrons, with $\Delta$ being the mean level spacing, to reverse the sign
of the persistent current\cite{riedel93}.

\subsubsection{Non-interacting electrons}\label{non-inter}
\paragraph{Average current}

The calculation of the average current\cite{cheung89} raises an
interesting problem of statistical physics.  In an experiment on many
rings, the number of electrons in each ring is fixed, whereas the 
chemical potential $\mu$ is not. We are thus dealing with the \emph{canonical}
ensemble\cite{bouchiat89,mont90}.  This point is very important, as it
has been shown that the persistent current calculated in the grand
canonical ensemble ($\mu$ fixed) is exponentially small, $I\approx
\exp(-L/2l_{e})$.

On the other hand, the calculation in the canonical ensemble can be 
related to the calculation in the grand canonical ensemble, which is much easier 
to perform\cite{imry97}. The canonical average persistent current is 
given by\cite{montambauxCours}:
\begin{equation}
\langle I_{N} \rangle =
-\left.\frac{\partial \langle
\mathcal{F}\rangle}{\partial\Phi}\right|_{N} = -\left.\frac{\partial
\langle\Omega\rangle}{\partial \Phi}\right|_{\mu}
\label{eq:Ican}
\end{equation}
where $N$ is the number of electrons, $\mathcal{F}$ the free
energy, $\Omega$ the grand potential and $\Phi$ the magnetic flux. 
With $\mu$ being the sample and flux dependent chemical potential,
one can then expand the expression (\ref{eq:Ican}) as a function of
$\delta\mu(\phi) = \mu(\phi) - \langle \mu \rangle$ where $\langle \mu
\rangle$ is flux independent:
\begin{eqnarray}
    -\left.\frac{\partial \langle\Omega\rangle}{\partial \phi}
    \right|_{\mu} & = & -\left.\frac{\partial
    \langle\Omega\rangle}{\partial\phi} \right|_{\langle\mu\rangle} -
    \delta\mu(\phi)\frac{\partial}{\partial\mu}\left.\frac{\partial\langle\Omega\rangle}{\partial\phi}\right|_{\langle\mu\rangle}
    \\ & = & -\left.\frac{\partial \langle\Omega\rangle}{\partial\phi}
    \right|_{\langle\mu\rangle} -
    \delta\mu(\phi)\frac{\partial}{\partial\phi}\left.\frac{\partial\langle\Omega\rangle}{\partial\mu}\right|_{\langle\mu\rangle}
    \label{eq:Ican3}
\end{eqnarray}
The first term is simply the grand canonical current, which is
exponentially small, and will be neglected.  The term
${\partial\langle\Omega\rangle}/{\partial\mu}$ corresponds to the 
number of
electrons. Using the relation $\delta\mu = -\delta
N\partial\mu/\partial N|_{\phi}$ we obtain: 
\begin{equation} 
\langle I_{N} \rangle =
-\left.\frac{\partial\mu}{\partial N}\right|_{\phi}\left\langle\delta
N\left.\frac{\partial
N}{\partial\phi}\right|_{\langle\mu\rangle}\right\rangle
\label{eq:Ican5}
\end{equation}
where $\left.\frac{\partial\mu}{\partial N}\right|_{\phi}$ is
the level spacing.  Finally, one
obtains\cite{altshuler91,schmid91,vonoppen91,akkermans91}:
\begin{equation}
\langle
I_{N} \rangle = -\frac{\Delta}{2}\frac{\partial}{\partial\phi}\langle
\delta N^2_{\mu}\rangle
    \label{eq:Ican6}
\end{equation}

The number of electrons is simply given by:
$N=\int_{-\varepsilon_{F}}^0 \rho (\varepsilon) d\varepsilon$.  The
fluctuation in the number of electrons is given by:
\begin{eqnarray*}
    \langle \delta N^2_{\mu}\rangle & = & \langle \left(N - \langle
    N\rangle\right)^2\rangle \\
    {} & = & \langle \int_{-\varepsilon_{F}}^0 \left(\rho
    (\varepsilon) - \rho_{0}\right)d\varepsilon
    \int_{-\varepsilon_{F}}^0 \left(\rho (\varepsilon^{\prime}) -
    \rho_{0}\right)d\varepsilon^{\prime}\rangle \\
    {} & = & \int_{-\varepsilon_{F}}^0\int_{-\varepsilon_{F}}^0
    \left(\langle \rho (\varepsilon)\rho (\varepsilon^{\prime})\rangle
    - \rho_{0}^2\right) d\varepsilon d\varepsilon^{\prime}\\
    {} & = & \int_{-\varepsilon_{F}}^0\int_{-\varepsilon_{F}}^0
    K(\varepsilon ,\varepsilon^{\prime}) d\varepsilon
    d\varepsilon^{\prime}
\end{eqnarray*}
where $K(\varepsilon ,\varepsilon^{\prime})$ is the two point
correlation function of the density of states.  The average current
is then given by:
\begin{equation}
  \langle I_{N} \rangle =
-\frac{\Delta}{2}\frac{\partial}{\partial\phi}
\int_{-\varepsilon_{F}}^0\int_{-\varepsilon_{F}}^0 K(\varepsilon
,\varepsilon^{\prime}) d\varepsilon d\varepsilon^{\prime}
    \label{eq:Ican7}
\end{equation}
It has been shown\cite{argaman93} that the spectral form factor
$\tilde{K}(t)$ can be related to the return probability to the origin
$P(t)=P(\vec{r},\vec{r},t)$:
\begin{equation}
\tilde{K}(t) = \frac{1}{4\pi ^2} t
P(t)
\label{eq:K(t)}
\end{equation}
which leads to: 
\begin{equation}
\langle I_{N} \rangle =
-\frac{\Delta}{4\pi ^2}\frac{\partial}{\partial\phi}\int_{0}^{\infty}
\frac{P(t)}{t} dt
\label{eq:Ican8}
\end{equation}
This return probability contains
two terms: the first one is flux independent and will be ignored. 
The interference term $P_{int}(t)$ can be expressed as a function of
the winding number of the different trajectories $m$:
\begin{equation}
P_{int}(t) = \sum_{m=-\infty}^{\infty} P_{m}(t)\cos \left(4\pi
m\frac{\phi}{\phi_{0}}\right)
\label{eq:Pint}
\end{equation}
Inserting this into
equation \ref{eq:Ican8}, one obtains:
\begin{equation}
\langle I_{N} \rangle =
\frac{2}{\pi}\frac{\Delta}{\phi_{0}}\sum_{m=1}^{\infty} m\sin
\left(4\pi m\frac{\phi}{\phi_{0}}\right) \int_{0}^{\infty}
\frac{P_{m}(t)}{t} dt
\label{eq:Ican9}
\end{equation}
Knowing the expression for $P_{m}(t)$: $ P_{m}(t)={1}/{\sqrt{4\pi
Dt}}\exp \left(-\frac{m^2L^2}{4Dt}\right)$, one 
finally obtains\cite{tablenum}:
\begin{equation}
\langle I_{N}\rangle = 
\frac{2}{\pi}
\frac{\Delta}{\phi_{0}} \sum_{m=1}^{+\infty}\sin\left(4\pi
m\frac{\phi}{\phi_{0}}\right)\exp\left(-m\frac{L}{l_{\phi}}\right)
\label{eq:Imoyen3D}
\end{equation}

This current has a periodicity of $\Phi_{0}/2$ and is paramagnetic for
small magnetic flux.  It should be stressed, however, that the 
amplitude of this current is of the order $\langle I_{N}\rangle \approx
\Delta/\phi_{0}$.  Taking a level spacing of $100\, \mu K$ (for a
metallic ring with a typical radius of one micrometer), one obtains a current 
in the range of the $pA$.  Such a current would certainly not be
measurable and is by far much lower than the experimentally observed 
value.

\paragraph{Typical current}

The typical current $I_{typ}$ is defined as the fluctuations around
the average current\cite{cheung89}:
\begin{equation}
I_{typ}^2 = \langle I^2\rangle
- \langle I\rangle^2\approx\sqrt{\langle I^2\rangle}
\label{eq:ItypDef}
\end{equation}
Starting from the expression for the current:
\begin{equation}
I=-\frac{\partial\mathcal{F}}{\partial\Phi}=\frac{\partial}{\partial
\phi}\int_{-E_{F}}^{0}\varepsilon\rho(\varepsilon,\Phi)d\varepsilon
\label{eq:DefI}
\end{equation}
one obtains:
\begin{eqnarray}
I_{typ}^{2} \approx \langle I^2\rangle & = &
\frac{\partial}{\partial\Phi}\frac{\partial}{\partial\Phi^{\prime}}\int_{-E_{F}}^{0}\varepsilon\varepsilon^{\prime}\langle
\rho(\varepsilon,\Phi)\rho(\varepsilon^{\prime},\Phi^{\prime})\rangle\,d\varepsilon
d\varepsilon^{\prime} \\ & = &
\frac{\partial}{\partial\Phi}\frac{\partial}{\partial\Phi^{\prime}}\int_{-E_{F}}^{0}\varepsilon\varepsilon^{\prime}K(\varepsilon
- \varepsilon^{\prime},\Phi,\Phi^{\prime})\,d\varepsilon
d\varepsilon^{\prime}
\end{eqnarray}

Performing a Fourier transform\footnote{We
have omitted the classical part of $P$ which does not depend on the
flux.} and using again the relation
\ref{eq:K(t)}, one obtains\cite{montambauxCours,argaman93}:
\begin{equation}
    I_{typ}^2 = \frac{1}{8\pi ^2}\frac{1}{\phi_{0}^2}
\int_{0}^{\infty}\frac{(t,\phi)}{t^3}dt
    \label{eq:ItypP}
\end{equation} where $P_{int}^{\prime\prime}$ denotes the second derivative of
$P_{int}$ with respect to $\Phi$.  Using the equation (\ref{eq:Pint}),
one finally obtains:
\begin{eqnarray}
 I_{typ}^2 =
 \frac{96}{\left(2\pi\right)^2}\left(\frac{E_{c}}{\phi_{0}}\right)^2\sum_{m=1}^{\infty}\frac{1}{m^3}
 \left[1+m\frac{L}{l_{\phi}}+\frac{1}{3}m^2\left(L/l_{\phi}\right)^2\right]\times\\
  \sin^2\left(2\pi 
m\frac{\phi}{\phi_{0}}\right)\exp\left(-m\frac{L}{l_{\phi}}\right)
\label{eq:ItypCarre}
\end{eqnarray}
Keeping only the first harmonic, and assuming $l_{\phi}\ll L$, we find
for the typical current\cite{tablenum}:
\begin{equation}
    I_{typ} \approx \frac{\sqrt{96}}{2\pi} \frac{E_{c}}{\phi_{0}}
    \approx 1.56 \frac{E_{c}}{\phi_{0}}
\label{eq:ItypFinal}
\end{equation}

This current is $\Phi_{0}$ periodic.  It should be noted that the
amplitude is again of order $E_{c}/\Phi_{0}$.  This result can be
rewritten as $I_{typ}\propto E_{c}/\Phi_{0}\propto e/\tau_{D}\propto
ev_{F}/L\cdot l_{e}/L$, where $\tau_{D}$ is the diffusion time. 
As derived in paragraph \ref{simple-model}, the typical current is
hence simply given by the time needed for an electron to perform one turn
around the ring.

\paragraph{Extensivity}

One important property of the typical current is the fact that its 
amplitude increases only as $\sqrt{N_{R}}$, where $N_{R}$ is the number of 
rings, since the typical current is given by the \emph{fluctuations} 
around the average value. 

On the contrary, the average current, such as any average value, grows
simply as $N_{R}$.  This has been extensively studied in the case of
conductance oscillations\cite{webb86}, but is also true for thermodynamics
properties.

\subsubsection{Interacting electrons}
Motivated by the first experimental observations, where a much larger 
amplitude of the persistent current has been obtained than 
theoretically predicted, electron-electron interaction have been 
recognized as an important contribution 
to the persistent current\cite{ambegaokar90}.  The calculation is 
made in the Hartree-Fock approximation, and one assumes a screened Coulomb
interaction\cite{aschcroft,kittel},
$U(\overrightarrow{r}-\overrightarrow{r}^{\prime}) =
U_{2D}\delta(\overrightarrow{r}-\overrightarrow{r}^{\prime})$, with
$U_{2D} = 2\pi e^2/q_{TF}$, $q_{TF}$ being the Thomas-Fermi wave
vector.  In the Hartree-Fock approximation, the total energy $E$
reads\cite{altshuler80a}:
\begin{equation}
E=E^{0}-\frac{U}{4}\frac{\partial}{\partial\Phi} \int
n^{2}(\vec{r})\,d\vec{r}
\end{equation} where $E^{0}$ is the total energy for the
non-interacting electrons.  Given that\footnote{The factor $2$ takes
into account for the spin.}
$n(\vec{r})=2\int_{0}^{\mu}\rho(\vec{r},\omega)\,d\omega$, one finds
for the interaction contribution to the average current:
\begin{equation}
    \langle
I_{ee}\rangle = \langle \frac{\partial E}{\partial\Phi}\rangle =
-{U}\frac{\partial}{\partial\Phi}\int\rho(\vec{r},\omega_{1})\rho(\vec{r},\omega_{2})\,d\vec{r}\,d\omega_{1}\,d\omega_{2}
\end{equation}

Again, this integral can be expressed as a function of $P(t)$:
\begin{equation}
\langle I_{ee}\rangle =
-\frac{U\phi_{0}}{\pi}\frac{\partial}{\partial\Phi}\int_{0}^{\infty}\frac{P(t,\Phi)}{t^2}\,dt
\end{equation}
Indexing by $m$ the winding number of the trajectories, one obtains
finally\cite{montambaux96}:
\begin{eqnarray}
    \langle I_{ee}\rangle =
    16\frac{U\rho_{0}}{2\pi}\lambda_{0}\frac{E_{c}}{\phi_{0}}\times \\
\sum_{m=1}^{+\infty}\frac{1}{m^2}\left[1+m\frac{L}{l_{\phi}}\right]\sin\left(4\pi   
m\frac{\phi}{\phi_{0}}\right)\exp\left(-m\frac{L}{l_{\phi}}\right)
    \label{eq:Iee}
\end{eqnarray}
with $\rho_{0}$ being the average density of states at the Fermi level
and $\lambda=U\rho_{0}$ the interaction coupling constant.  In the simple
limit $l_{\phi}\ll L$, and considering only the first harmonic, one
finds an average current of the order $E_{c}/\Phi_{0}$, a result
obtained in the simple model of the paragraph \ref{simple-model}.  It
should be noted that this current is much larger than the
non-interacting current calculated in the paragraph \ref{non-inter}. 
Another interesting point is that the prefactor is proportional to the
interaction parameter $U$: this implies that the sign of the average
current depends on the attractive or repulsive nature of the
interaction.  Finally, it should be stressed that this result is
independent of the statistical ensemble: coulombian interactions fixes
locally the electron density\cite{schmid91,ambegaokar90}, leading
to this insensitivity to the statistical ensemble.
Calculations including exact coulombian interactions lead to somehow
unclear results\cite{muller93,bouzerar94,abraham93,kato94,bouzerar95}.

We should also mention that the fluctuations of the persistent
current (the typical current) are much larger than its average value
(even when including the interaction term): $\sqrt{\langle I^2
\rangle}\gg \langle I \rangle$.  However, the typical current for
$N_{R}$ rings varies as $\sqrt{N_{R}}$, whereas the average current
varies as $N_{R}$: for few (or single) rings experiments, the
$\Phi_{0}$ periodic typical current dominates, whereas for a large
number of rings, the signal is dominated by the $\Phi_{0}/2$ periodic
average current.

Finally, it should be stressed that the calculation of the typical
current is made \emph{only} for non-interacting electrons.  Attempts
to include coulombian interactions\cite{ramin95,eckern92,smith92} are
more difficult to perform and interpret.

\section{Experimental results}

\subsection{Orders of magnitude}\label{magnitude}
Due to the experimental difficulty, only few experimental 
studies on persistent currents are available. 
In the following we will give a review of these experiments. 
There are two distinct sets of experiments: first the \emph{many ring 
experiments} which have been carried out on a very large number 
of rings. Secondly, the \emph{single ring experiments}. 
Both kind of experiments have been performed on metals 
and semiconductor heterojunctions.
More recently, experiments have been carried out on a small number of 
rings.

Let us recall briefly the order of magnitude for the typical and
average current.  The typical current, for $N_{R}$ rings, is given by:
\begin{equation}
    I_{typ}=\sqrt{\langle I^2
\rangle}=\frac{\sqrt{96}}{2\pi}\,\frac{ev_{F}}{L}\,\frac{l_{e}}{L}\,
\sqrt{N_{R}}\approx1.56\,\frac{ev_{F}}{L}\,\frac{l_{e}}{L}\,\sqrt{N_{R}}
\end{equation}
whereas the average current is given by:
\begin{equation}
    \langle I\rangle =
\frac{16}{2\pi}\,\lambda\,\frac{E_{c}}{\phi_{0}}\,N_{R} 
\end{equation} 
The coupling constant $\lambda$, when taking into account all the 
orders
of the interactions, is typically of the order $10^{-1}$.  This gives
for the average current: 
\begin{equation}
    \langle I\rangle\approx
0.25\,\frac{ev_{F}}{L}\,\frac{l_{e}}{L}\,N_{R} 
\end{equation} 
The average
current, even when taking into account coulombian interactions, is one
order of magnitude lower than the typical current.

In a metal, the Fermi velocity is of the order of $10^7\,m\,s^{-1}$
whereas for semiconductors, it is typically $10^5\,m\,s^{-1}$ 
The elastic mean free path in a metal is typically $20\, nm$ and about 
 $10\,\mu m$ in a heterojunction. For $N_{R}$ rings
of radius $2\,\mu m$, one obtains:
\begin{itemize}
\item For metals:
\begin{equation}I_{typ}\approx 0.3\,\sqrt{N_{R}}\,[nA]\approx
430\,\sqrt{N_{R}}\,[\mu_{B}]
\end{equation}
\begin{equation}
\langle I\rangle\approx
0.05\,N_{R}\,[nA]\approx 70\,N_{R}\,[\mu_{B}]
\end{equation}
\begin{equation}
E_{c}\approx 60\,mK
\end{equation}
\item For semiconductor heterojunctions:
\begin{equation} I_{typ}\approx
2\,\sqrt{N_{R}}\,[nA]\approx 2700\,\sqrt{N_{R}}\,[\mu_{B}]
\end{equation}
\begin{equation}
\langle I\rangle\approx 0.5\,N_{R}\,[nA]\approx 670\,N_{R}\,[\mu_{B}]
\end{equation}
\begin{equation}E_{c}\approx 380\,mK\end{equation}
\end{itemize}
For single or few rings experiments, the signal is dominated by the
typical current, whereas for many rings experiments, it is the average
current which is measured.  However, in all cases, the signal to be
measured is rather small, and such experiments are always an
experimental challenge.

In all the experiments performed up to now, the signal detected is the
magnetic flux generated by the persistent currents.  Basically, two
different techniques have been employed: first the dc \squid, either a
macroscopic (standard) one\cite{clarke76} or an on-chip
micro-\squid\cite{chapelier91}.  The second one involves an RF
resonator that allows to detect at the same time both the magnetic
flux generated and the ``conductivity'' of isolated rings.

\subsection{Many rings experiments}
In many rings experiments, at least when the number of rings is very
large, the measured physical quantity is the average current, as it
grows like the number of rings.  These experiments are also easier to
perform, as one deals with a ``macroscopic'' object, and hence the
detector is simpler to design.

\subsubsection{Metallic rings}
\begin{figure}[tbp]
\centerline{\includegraphics*[width=8cm]{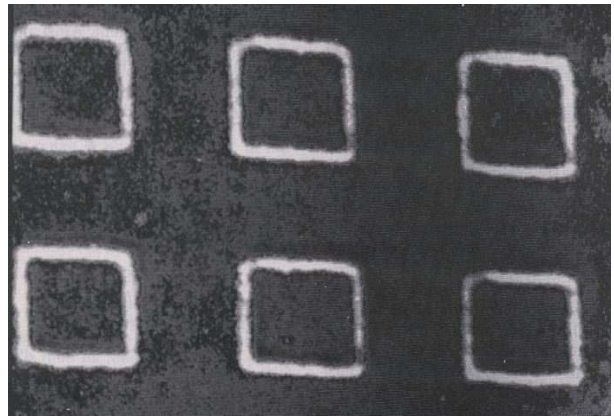}}
\caption{\textsc{Sem} picture of a part of the sample used 
in the experiment of ref.\cite{levy90}. It consists 
of an array of $10^7$ copper squares, of perimeter $2.2\,\mu m$. After 
ref.\cite{levy90}.}
\label{Levy01}
\end{figure}

The first experimental observation of the existence of persistent 
currents 
has been performed by L\'evy and coworkers on a network of $10^7$ copper
rings\cite{levy90} as shown in figure \ref{Levy01}.  In this 
experiment, the rings were squares of perimeter $2.2\,\mu m$, 
which gives $\Phi_{0}\approx 130\,G$ \footnote{The sample specific 
parameters in this experiments were: $l_{e}=20\,nm$; $E_{c}$ = $80\,mK$} 
and the phase coherence length was much larger than the perimeter of the rings.

The signal is detected using a commercial dc \squid.  It is
crucial to eliminate the contribution due to magnetic impurities from
the signal of the rings.  For that purpose, the authors used the
non-linearity of the signal coming from the persistent currents: the
magnetic field is modulated at low frequency and the signal is
detected as the second and third harmonic of the magnetic response. 
The procedure is repeated at several values of the magnetic field.

\begin{figure}[tbp]
\centerline{\includegraphics*[width=8cm]{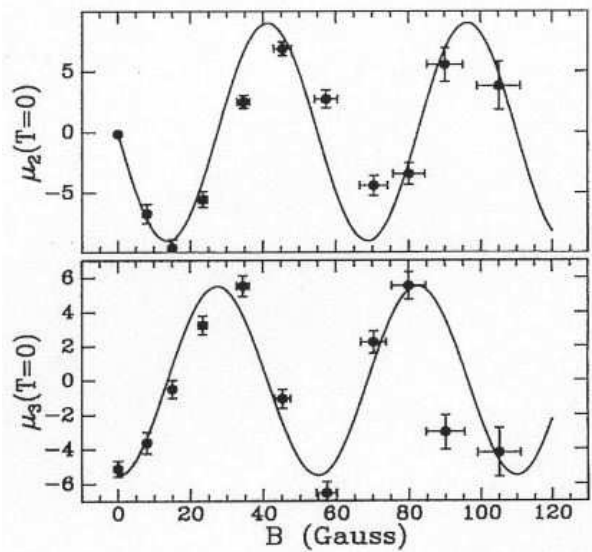}}
\caption{Dependence of the second and third harmonic of the response 
of the \squid~as a function of the magnetic field. In this 
experiment\cite{levy90}, $\Phi_{0}$ corresponds to $130\,G$. Both 
harmonics show clear oscillations as a function of the magnetic 
field, with a periodicity $\Phi_{0}/2$. After 
ref.\cite{levy90}.}
\label{Levy02}
\end{figure}

The experimental data are reported on figure \ref{Levy02}.  The signal
displays clear oscillations as a function of the magnetic field, with 
periodicity
$\Phi_{0}/2$.  The amplitude of the persistent current, deduced from
the magnetic response, is $0.4\,nA$ per ring, corresponding to $3\cdot
10^{-3}\,ev_{F}/L$ per ring.  This result, although somehow larger
than predicted, is in relatively good agreement with theory  taking
into account electron electron interactions.

In this experiment, the determination of the sign of the magnetic
response relies on some assumptions for the data processing.  In the
paper, the authors stated a \emph{diamagnetic} response at zero 
field. 
This result is quite surprising as it would correspond to an
\emph{attractive} interaction, which is quite unlikely in a metal like
copper. On the other hand, this sign has been confirmed by a recent 
experiment on silver rings\cite{deblock02}.

\subsubsection{Semiconductor rings}
Another experiment has been performed on a large number of rings in a 
semiconductor heterojunction\cite{reulet95}. 
The rings were $10^{5}$ squares of mean perimeter $8\,\mu 
m$ (corresponding to $\Phi_{0}\approx 10\,G$)\footnote{the sample 
specific parameters in this experiment were:$l_e$=$3\,\mu m$; $E_{c}$ 
= $200\,mK$, $l_{\phi}$ = $8\,\mu m$.}. In such semiconductor rings, 
the level spacing $\Delta$
is of the order of $25\,mK$, much higher than in metallic rings where
it is of the order of $\approx 10\,\mu K$.  
In this experiment, the experimental technique to detect the 
persistent current is somewhat different than the technique used in 
the experiment by L\'evy \textit{et al.}. Instead of measuring the dc 
magnetic response of the rings, the authors study the ac response of 
the rings to an rf excitation.  Using this technique, they measure 
the ac complex
conductance of the rings, from which they deduce the persistent 
currents.
\begin{figure}[tbp]
\centerline{\includegraphics*[width=8cm]{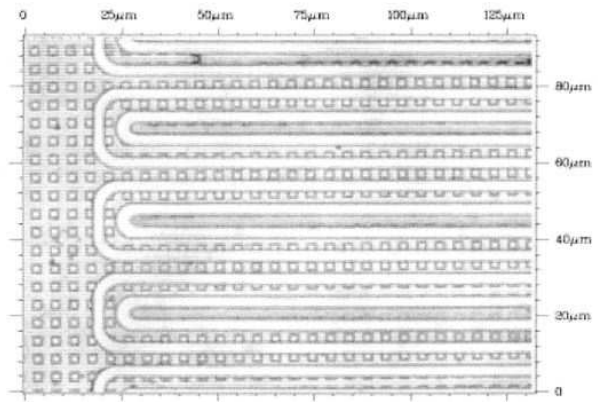}}
\caption{Optical photograph of the sample used in the experiment of\cite{reulet95}. It consists of an array of $10^5$ GaAs/AlGaAs rings. 
On the top of them, one can 
see the niobium meander stripe-line used as the resonator. Reprinted 
from ref.\cite{reulet95}.}
\label{Reulet01}
\end{figure}

The quantity measured in this experiment is the magnetic
susceptibility of the rings,
$\chi(\omega)=\chi^{\prime}(\omega)+\imath\chi^{\prime\prime}(\omega)$.
The complex ac conductance of the rings is then deduced by
$\chi(\omega)\propto \imath\omega G(\omega)$.  At low
frequency\footnote{In this experiment, the characteristic frequency is
given by the inverse of the inelastic mean free time
$\tau_{in}^{-1}$.}, the imaginary part of $G(\omega)$ is just
proportional to the derivative of the persistent current with respect
to the flux.

The magnetic susceptibility is measured using a resonating
technique.  The resonator consists of a meander strip-line on top of 
which the
rings are deposited.  The meander, open at both ends, is made of
$20\,cm$ of superconducting niobium.  The fundamental frequency of the
resonator is $380\,MHz$.  The shift in the resonance frequency and
the variations of the quality factor are proportional to the
imaginary and real parts of the ac complex conductance of the rings.

\begin{figure}[tbp]
\centerline{\includegraphics*[width=8cm]{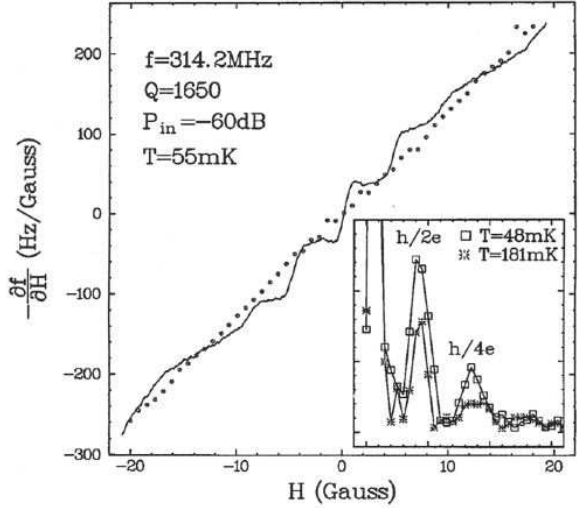}}
\caption{Derivative of the resonance frequency of the resonator used 
in ref.\cite{reulet95}. The linear background (dotted line) is due to 
the diamagnetism of niobium. Superimposed on this, one 
clearly sees the  $h/2e$ periodic signal due to the persistent currents in the 
rings(solid line). The inset shows the Fourier transform of the signal at two 
different temperatures. Reprinted 
from ref.\cite{reulet95}.}
\label{Reulet02}
\end{figure}

The experimental conductance (see figure \ref{Reulet02}) shows $h/2e$
oscillations, as expected for experiments on many rings.  However, the
amplitude found for the persistent current, on the order of $1.5\,nA$ per ring,
is almost an order of magnitude larger than predicted.  More 
surprising, the measured signal implies a \emph{diamagnetic} zero field persistent
current, \textit{i. e.} an \emph{attractive} interaction between the
electrons.  Again, such an attractive interaction is very unlikely in
this two-dimensional electron gas.

It should be noted, however, that in such an experiment, the frequency
is quite close to the level spacing.  This may affect the response of
the rings and makes a direct comparison with the experiment by L\'evy
\textit{et al.} somewhat difficult.

\subsection{Single rings experiments}
Single ring experiments are a true experimental challenge.  In such
experiments, it is the \emph{typical} current which is detected, as it
is roughly one order of magnitude larger than the average current for
one ring.  It should be stressed, however, that as the average current
is an extensive quantity in contrary to the typical current that
varies like $\sqrt{N_{R}}$, the signal to be detected in a many rings
experiment is orders of magnitude larger than the signal to be
detected in a single ring experiment.
\subsubsection{Metallic rings}
The first single ring experiment has been performed by Chandrasekhar
\textit{et al.}\cite{chandrasekhar91} on a single gold ring.  In this
experiment, three different samples have been measured: two were rings
of diameter $2.4\,\mu m$ and $4.0\,\mu m$, and the third one was a rectangle of
dimensions $1.4\,\mu m\times 2.6\,\mu m$\footnote{$l_{\phi}$ = $12\,\mu m$; $l_{e}\approx 70\,nm$ in this
experiment.}.

\begin{figure}[tbp]
\centerline{\includegraphics*[width=8cm]{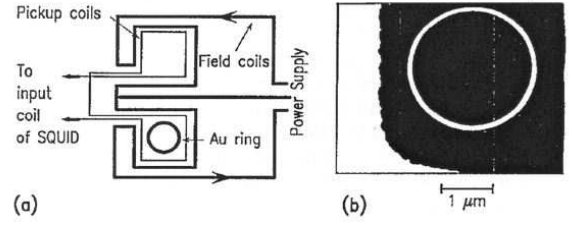}}
\caption{a) Schematic picture of the sample used for the experiment 
of ref.\cite{chandrasekhar91}, displaying the counter wound niobium 
pick-up loop, the field coil and the gold ring. b) \textsc{Sem} 
picture of the ring. The white part is a corner of the pick-up loop. Reprinted 
from ref.\cite{chandrasekhar91}.}
\label{Chandrasekhar01}
\end{figure}

The experimental setup consists of a home made miniature dc \squid. 
The \squid, itself has a sensitivity of $6\cdot 10^{-8} \Phi_{0}$.  The
pick-up loop consists of a counter wound niobium loop in order to
minimize the sensitivity to the static background field.  To maximize
the coupling between the pick-up coil and the sample, both have been
fabricated on the same chip, and the coil is deposited around the
gold ring (see figure \ref{Chandrasekhar01}).  Moreover, the field 
coil consists of a niobium line deposited around the ring.

\begin{figure}[tbp]
\centerline{\includegraphics*[width=8cm]{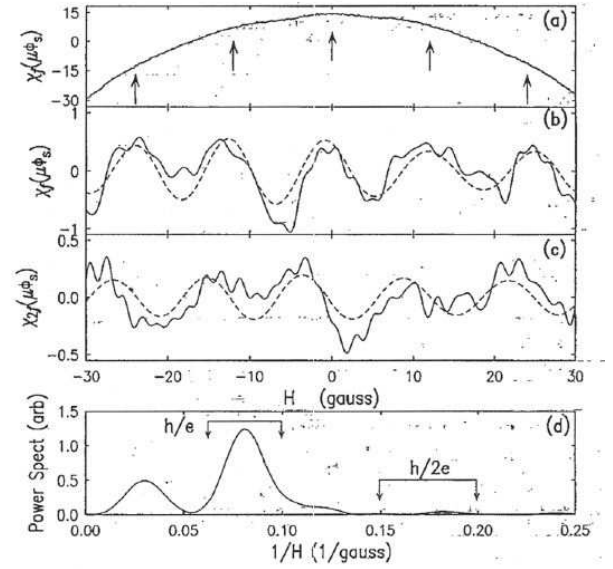}}
\caption{a) First harmonic of the response of the dc \squid~as a 
function of the magnetic field. b) Same data after subtraction of a 
quadratic background. c) Second ($2\,f$) harmonic of the response of 
the dc \squid~after subtraction of a quadratic background. d) Power 
spectrum of the data displayed in b). One clearly observes a peak at 
$h/e$ frequency. Reprinted from ref.\cite{chandrasekhar91}.}
\label{Chandrasekhar02}
\end{figure}

In this experiment, the authors detect the modulation of the flux
measured by the \squid~as a function of the magnetic field, which is
swept over a few $\Phi_{0}$. The magnetic field is modulated at 
low frequency ($\approx 4\,Hz$), and the signal detected at $f$ and 
$2\,f$. The background signal is subtracted numerically using a 
quadratic form and the amplitude is extracted from the Fourier 
transform (power spectrum) of the data as a
function of the magnetic field (see figure \ref{Chandrasekhar02}).

As a result, the author found a persistent current with $\Phi_{0}$
periodicity and an amplitude of $3\pm 2\,nA$, $30\pm 15\,nA$ and 
$6\pm 2\,nA$ for the three samples investigated, whereas the theoretical 
values are $0.09\,nA$, $0.27\,nA$ and $0.25\,nA$ respectively.  Obviously, the measured
signal is $30$ to $150$ times larger than expected.  Different
arguments have been invoked to explain this discrepancy.  It
should be noted, however, that the observed signal is of the order $ev_{F}/L$,
\textit{i. e.} the signal one should find for a ballistic ring
($l_{e}\approx L$).  On the other hand, it is very unlikely that gold rings
behave as ballistic rings, and the theoretical explanation of this 
experimental observation remains an open question.  

The sign of the persistent current is quite difficult to determine in
such a single ring experiment.  The authors
claim that the samples studied showed a paramagnetic signal. 
A clear statement, however, as stressed by the authors is difficult 
due to the few samples measured and due to the extreme experimental 
difficulty.

\subsubsection{Semiconductor rings}
Another experiment has been performed by Mailly and coworkers on a 
single, isolated ring, etched into a semiconductor 
heterojunction\cite{mailly93}.  In this
case, the signal is expected to be larger than in metallic rings as
the elastic mean free path is much larger compared to the latter case.

The ring has been etched into a two dimensional electron gas at the
interface of a GaAs-GaAlAs heterojunction.  The mean 
perimeter in this sample is on the order of $6\,\mu m$ which 
corresponds to $\Phi_{0}\approx 10\,G$)\footnote{$l_{e}\approx 
10\,\mu m$;$l_{\Phi}\approx 25\,\mu m$; $E_{c}\succ 1\,K$ in this 
experiment}.  

\begin{figure}[tbp]
\centerline{\includegraphics*[width=8cm]{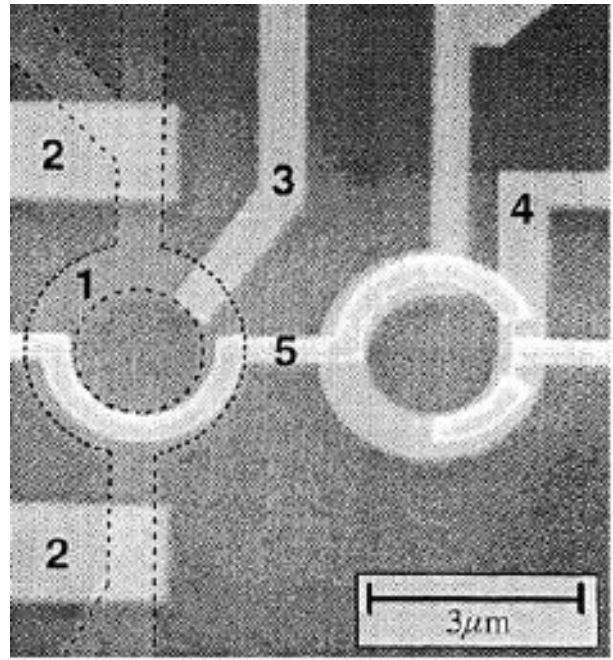}}
\caption{\textsc{Sem} picture of the sample used in 
ref.\cite{mailly93}. 1)~GaAs/GaAlAs ring (dashed line). 2) and 3) gold 
gate used to isolate the sample from ohmic contacts and to suppress 
the signal. 4) Gold calibration loop. 5) First level of the 
micro-\squid~gradiometer containing the two Dayem microbridges on the 
right. The picture has been taken before the evaporation of the 
second level of the miro-\squid~gradiometer. Reprinted from ref.\cite{mailly93}.}
\label{Mailly01}
\end{figure}

An important advantage of semiconductors is the possibility of using
gates on the sample.  This allows to modify \textit{in situ} the
geometry of the sample simply by applying a dc voltage to the gates. 
In this experiment, the authors used two different gates (see figure
\ref{Mailly01}): the first is used to separate the ring from the
reservoirs.  The presence of these ohmic contacts allows to measure at
the same time the conductance and the persistent current oscillations,
and hence to check the electronic temperature and the coherence of the
electrons in the ring.  The second gate is evaporated on top of
one arm of the ring.  By polarizing this gate (``open'' ring), one can
suppress all the interference effects in the ring, both the
Aharonov-Bohm oscillations and the persistent currents.  This allows
to perform a ``zero'' measurement, equivalent to measuring the
\squid~with no ring.  The advantage is that this can be made on the
same sample.  Moreover, the subtraction of the signal obtained with
the ring ``closed'' and ``open'' allows to experimentally suppress the
background signal of the detector.

In this experiment, a sophisticated on-chip micro-\squid~technique has been 
employed. With such a design, no pick-up coil is needed: the \squid~itself is
deposited exactly on the top of the ring.  This has two major
advantages: first, the absence of a pick-up coil reduces the
inductance of the setup.  Secondly, and most important, in such
a geometry, the coupling between the ring and the \squid~is basically
optimal, as the \squid~has exactly the same shape as the ring.  The
\squid~is actually designed as a gradiometer, consisting of two 
counterwound loops
in order to compensate the externally applied static magnetic field 
(see figure \ref{Mailly01}). 
The two Josephson junctions are made using Dayem microbridges, 
evaporated at the same time as the
second level of the gradiometer. For a detailed description of the 
miro-\squid~gradiometer  technique, we refer the reader to 
the reference\cite{rabaud_ms+s2002}.

The measurement consists in sweeping the magnetic field over several
$\Phi_{0}$ and recording the critical current of the \squid.  This
is made successively for the ``closed'' and ``open'' ring.  The signal
is then obtained by taking the Fourier transform of the difference
between the two measurements.  The noise is evaluated at the same time
by taking the Fourier transform of the difference between two 
``closed'' or ``open'' ring measurements.

\begin{figure}[tbp]
\centerline{\includegraphics*[width=8cm]{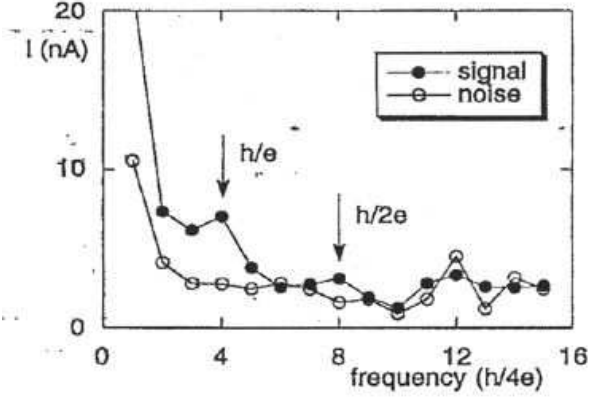}}
\caption{Fourier spectrum of the magnetization of the ring of 
ref.\cite{mailly93} in units 
of $nA$. The arrows indicate the $h/e$ and $h/2e$ frequency. Open dot 
is the experimental noise. One clearly observes the signal at the 
$h/e$ frequency corresponding to a persistent current of $4\pm 2\,nA$ in 
the ring. Reprinted from ref.\cite{mailly93}.}
\label{Mailly02}
\end{figure}

In the Fourier spectrum (see figure \ref{Mailly02}), a clear peak is 
observed at the $\Phi_{0}$ frequency corresponding to a value of $4\pm 
2\,nA$ for the persistent current amplitude, in good agreement with 
the theoretical prediction $ev_{F}/L$.  No measurable signal was 
observed at the $\Phi_{0}/2$ frequency, as expected when comparing 
the theoretical signal and the noise level of the experiment.  
The sign of the persistent current, on the other hand, was impossible 
to determine in a reliable way.

This experiment proves that in the case of very weak disorder and
small number of channels, standard theory gives a correct description
of the persistent current amplitude.  Moreover, in such samples,
electron-electron interactions are much enhanced due to the low
electron density. This suggests that these interactions are unlikely 
to strongly enhance the amplitude of the persistent currents.

\subsection{Few rings experiments}
More recently, two experiments have also been performed on ensemble
of few rings, either metallic or semiconductor.  In this case, the
small number of rings (typically ten rings) allows to check the theory
concerning the ensemble averaging, and should allow to observe both
the $h/e$ and $h/2e$ components of the persistent current.  Moreover,
in the experiment on semiconductor rings, the authors were able to
check the effect of a connection (ohmic contacts) between the rings.

\subsubsection{Metallic rings}
\begin{figure}[tbp]
\centerline{\includegraphics*[width=8cm]{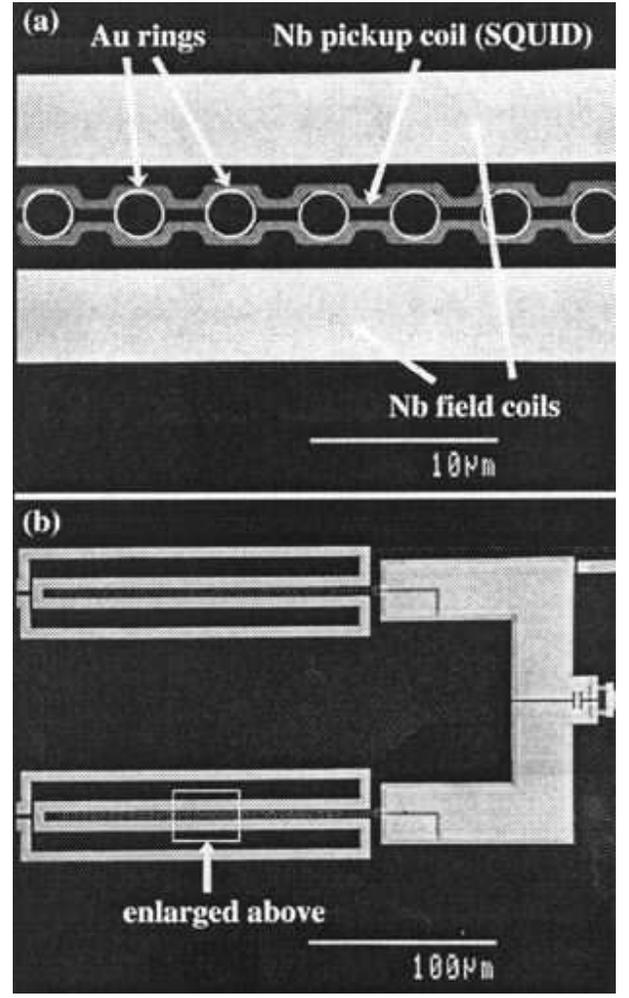}}
\caption{\textsc{Sem} picture of the sample used in the experiment of 
ref.\cite{jariwala00}. a) Close-up view showing the gold rings, the 
niobium pick-up coil and the niobium field coil. b) Larger view 
showing the entire gradiometer. Reprinted from ref.\cite{jariwala00}.}
\label{Jariwala01}
\end{figure}

In this experiment, Jariwala et al.\cite{jariwala00} used a similar
experimental set-up as for the experiment on the single gold ring.  The
sample (see figure \ref{Jariwala01}) consists of a line of $30$ 
isolated gold rings  of radius $1.3\,\mu m$ (perimeter $8\,\mu m$) 
corresponding to a flux period of $\Phi_{0}\approx
8\,G$\footnote{$l_{e}\approx 87\, nm$; $l_{\Phi}\approx 16\,\mu m$; 
$E_{c}\approx 7\,mK$ in this experiment.}.

To extract the persistent current signal from the background signal,
the magnetic field is modulated at low frequency (typically $\approx
2\,Hz$) and detect at the first, second and third harmonic of the
response of the \squid.

\begin{figure}[tbp]
\centerline{\includegraphics*[width=8cm]{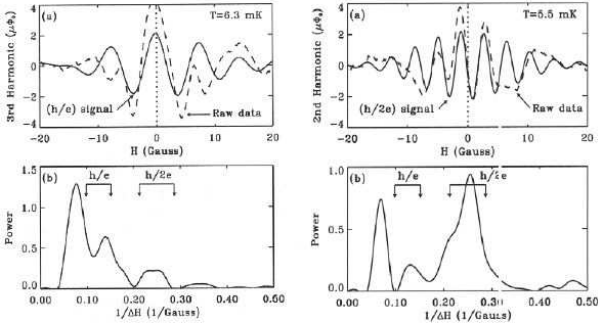}}
\caption{Magnetic response of the array of $30$ gold rings of the 
experiment of ref.\cite{jariwala00}. Left panel: a) Raw data (dashed 
line) and $h/e$ contribution (solid line) extracted from the Fourier 
spectrum displayed in b). Right panel: a) Raw data (dashed 
line) and $h/2e$ contribution (solid line) extracted from the Fourier 
spectrum displayed in b). Reprinted from ref.\cite{jariwala00}.}
\label{Jariwala02}
\end{figure}

In this experiment, both the $h/e$ and $h/2e$ components were detected
(see figure \ref{Jariwala02}).  For the $h/e$ component, the authors
found a current of $I_{typ}=0.35\,nA=2.3\,E_{c}/\Phi_{0}$ per ring, in
good agreement with theoretical predictions, taking into account e-e
interactions. To obtain this result, the authors divided the total
signal by $\sqrt{N_{R}}$ to account for the random sign of the
persistent current.  This suggests that the amplitude of the
persistent current measured in the single ring
experiment\cite{chandrasekhar91} is somewhat overestimated.

The $h/2e$ component was found to be $\langle I_{N} \rangle =
0.06\,nA$ per ring, corresponding to $0.44\,E_{c}/\Phi_{0}$.  This
result is in line with the results found in the previous experiment of
L\'evy \textit{et al.} on copper rings, and only a factor of $2$ larger than the
theoretical predictions when taking into account electron-electron interactions. 

The sign of the persistent current on the other hand is much more
surprising.  In this experiment the sign of the average current is
\emph{diamagnetic}.  Although this has been seen in previous
experiments on many rings\cite{levy90,deblock02}, in this work the
determination of the sign is unambiguous.  As we have seen, such a
diamagnetic response is quite unlikely as it corresponds to an
\emph{attractive} interactions between the electrons.  Clearly, such an
discrepancy between experiment and theory may be attributed to an
unexplored physical phenomena that modifies the ground state of the
electron gas.  The authors of this experiment explain their result in
the light of a recent theory on zero temperature dephasing in
metals\cite{kravtsov00}.  However, as the status of such theories is
still quite controversial, we will not go further into this point.

Finally, it should be noted that in such an experiment on $30$ rings,
both the average current, that grows like $N_{R}$, and the typical
current, that grows like $\sqrt{N_{R}}$, have the same amplitude. 
This proves that $30$ rings are not enough for ensemble averaging, and
the exact variation of $I_{typ}$ and $\langle I_{N} \rangle$ with the
number of rings remains experimentally an open question.

\subsubsection{Semiconductor rings}
\begin{figure}[tbp]
\centerline{\includegraphics*[width=8cm]{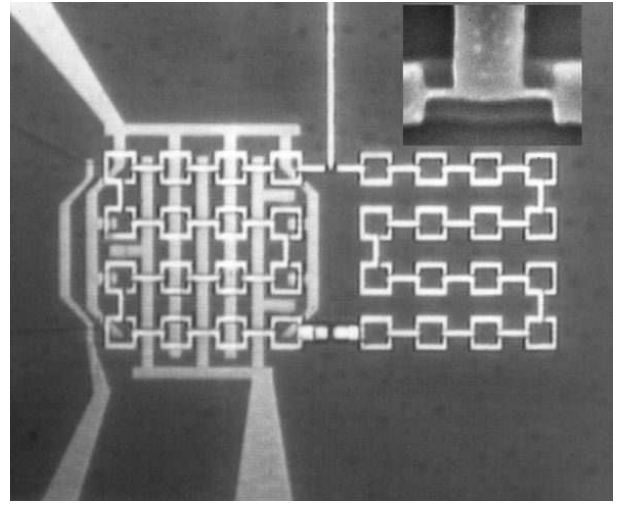}}
\caption{optical photograph of the sample used for the experiment of 
ref.\cite{rabaud01}. The three metallic gates and the aluminium 
micro-\squid~gradiometer are clearly visible. Inset shows a 
\textsc{Sem} picture of the two Dayem microbridges used as Josephson 
junctions for the \textsc{squid}. After ref.\cite{rabaud01}.}
\label{Rabaud01}
\end{figure}

Another experiment has been performed on a small number of 
semiconductor rings by
Rabaud and coworkers\cite{rabaud01}.  This experiment has been 
performed on two arrays
of $4$ and $16$ rings (actually squares) etched into a two 
dimensional electron gas at the interface of a GaAs GaAlAs 
heterojunction. 
The squares were $3\,\mu m\times 3\,\mu m$, perimeter
$12\,\mu m$, corresponding to $\Phi_{0}\approx 
5\,G$)\footnote{$l_{e}\approx 10\,\mu m$; $l_{\Phi}\approx 25\,\mu 
m$; $E_{c}\approx 500\,mK$ in this experiment}. 
Using an original set-up containing three different metallic gates (see 
figure \ref{Rabaud01}), Rabaud et al. have been able to measure in 
the same experiment (same cooling down
run) both, the signal of connected and isolated rings.  The original 
set-up also permits to suppress the signal via a gate and check the 
``zero'' of the detector.  This allows to perform an ``in situ''
subtraction of the background.  Such a technique has the advantage to 
measure at the same time both the signal and the noise in order to 
have an unambiguous determination of the magnetic signal.

\begin{figure}[tbp]
\centerline{\includegraphics*[width=8cm]{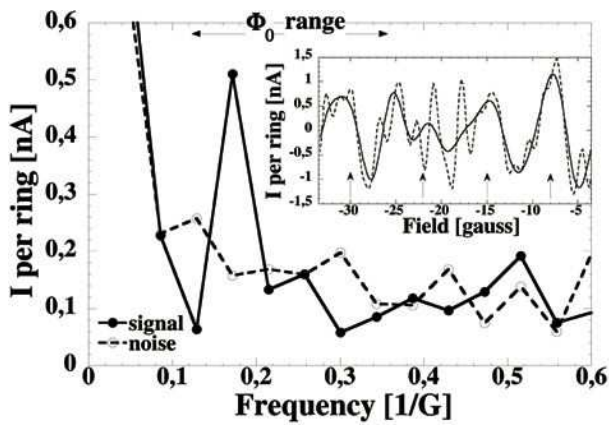}}
\caption{Power spectrum of the magnetization due to the persistent in 
a line of $16$ connected rings in units of $nA$ per ring. The arrow 
indicates the $h/e$ frequency window. Open symbols are the 
experimental noise. One clearly sees a peak in in the ``signal'' 
curve, absent from the ``noise'' curve. Inset shows the raw data 
(dashed line) after substraction of the background and after 
bandpassing the signal over the $h/e$ 
frequency range (solid line). After ref.\cite{rabaud01}.}
\label{Rabaud02}
\end{figure}

In this experiment, the authors measured a clear $h/e$ periodic
signal, of amplitude $2.0\pm 0.3\,nA$ per ring for the $4$ rings
sample, and $0.35\pm 0.07\,nA$ per ring for the $16$ rings sample (see
figure \ref{Rabaud02}), to be compared with the theoretical values
$2.18\,nA$ per ring for the $4$ rings sample and $1.09\,nA$ per ring
for the $16$ rings sample.  The experimental results are in relatively
good agreement with the theoretical values.  The discrepancy observed
for the $16$ rings sample may be attributed to an overestimation of
the elastic mean free path, which is determined on wires fabricated
from the same wafer of GaAs/AlGaAs heterojunction.  However, the
complete lithographic process, quite complicated in this experiment,
may affect $l_{e}$, mainly because of the roughness of the edges after
etching.

In this work, the authors were also able to measure the persistent
currents in the same array of rings, but this time with an ohmic
connection between the rings.  Measurements on both, isolated and
connected rings can be made basically at the same time, by simply applying a dc
voltage on the gates on the top of the arms connecting the rings.  
The
purpose of this experiment was to measure the persistent current in a
sample much larger than $l_{\phi}$, \textit{i. e.} a macroscopic sample
from the quantum physic point of view.  This work was stimulated by
theoretical models that calculated persistent currents in arrays of
rings, showing that they do not vanish, but are only reduced by
some geometrical factor\cite{pascaud99}.

The lines of $4$ and $16$ rings used in the experiment were
respectively $\approx 60\mu m$ and $\approx 250\mu m$, both much
larger than the phase coherence length.  In that sense, these line of
rings are macroscopic objects.  The authors found a current of
amplitude $1.7\pm 0.3\,nA$ per ring for the $4$ rings sample, and
$0.40\pm 0.08\,nA$ per ring for the $16$ rings sample, whereas the
theoretical values, calculated in ref.\cite{pascaud99}, were 
respectively $1.25\,nA$
per ring and $0.62\,nA$ per ring.  There is obviously a discrepancy
between experimental and theoretical values.  However, it should be
noted that the theoretical model has been developed for diffusive
(metallic) rings, which is certainly not the case in heterojunction
rings.  Moreover, coulomb interactions are not taken into account
for the typical current; in heterojunctions, the low electronic
density enhances strongly the interactions.  

The key result of this experiment is the fact that the
ratio between the amplitude of the persistent currents observed in
connected and isolated rings is of the order of one for both samples. 
This shows that persistent currents are basically unaffected by the
connection between the rings. This suggests that even in
a macroscopic sample, there should be a reminiscence of the quantum nature of
electrons.

Finally, it is interesting to compare this experiment with the
experiment performed on $30$ metallic rings.  In the experiment on
semiconductor rings, no observable signal was detected at the $h/2e$
frequency for both samples.  At least for the $16$ rings sample, this
result is quite surprising, as the average current grows linearly with
the number of rings. As a comparison, in the experiment on metallic
rings, the signals at $h/e$ and $h/2e$ for $30$ rings were of similar
amplitude. Again, this shows that the ensemble averaging, in the case
of persistent currents, is still not fully understood.


\section{Conclusion}

Persistent currents are certainly one of the most spectacular
manifestation of the quantum coherence of the electrons in a
mesoscopic system: it manifests as a permanent, non dissipative
current flowing around a normal, non superconducting ring.  The
amplitude of this current is of the order of a nanoAmpere, whereas the
resistance of the ring can be of the order of a kiloOhms, \textit{e. g.} 
for the case of semiconductor rings.

Although heavily controversial at the beginning, the existence of such
currents is well established, from both, theoretical as well as
experimental point of view.  However, many questions remain open, and
experimental results point out the lack of a deep understanding of
this phenomenon.

First the amplitude experimentally observed seems different from the
theoretical predictions.  Most of the experimental results are about
an order of magnitude larger than the theoretical predictions. 
However, it must be stressed that all these experiments are very
difficult, as they deal with the measurement of very small magnetic
signals.  From this point of view, and taking into account the
different approximations in the theoretical models, it seems difficult
in the absence of new experimental results to draw a definitive
conclusion concerning the validity of the theoretical predictions on
the amplitude of the persistent currents.

More surprising is the sign observed in the many rings experiments. 
As we have seen, the sign of the zero field magnetic response due to
the average persistent currents are directly related to the sign of
the interaction between electrons.  In at least two experiments on
many rings, both metallic or semiconductor, the sign was found to be
diamagnetic, whereas in the first experiment on copper rings, there
were indications that it was also diamagnetic.  This result is quite
intriguing, as it should correspond to an attractive interaction
between electrons.  Such an attractive interaction is very unlikely in
``standard'' metals like copper or gold, or even in GaAs/GaAlAs
heteronjunctions.  Clearly, there are many open questions in the
description of the average persistent current of interacting
electrons.

Another interesting point is the change of persistent currents when the
sample evolves from a true mesoscopic sample to a macroscopic sample.  
Only one such experiment has been carried out up to now, and the result found
is that persistent currents are not significantly modified when the
size of the sample increases. These results suggest that persistent
currents should be observable in a macroscopic objet; by extension,
in the spirit of the evolution from Aharonov-Bohm oscillations to
weak localisation, one may think about observing the zero field
magnetic response of a standard two dimensional metal.

Finally, there are natural extensions of this problem of equilibrium
properties of mesoscopic conductors, that have been largely 
unexplored. 
An interesting problem is a ballistic dot of different
shape, \textit{ie} quantum billiards.  In this case, the properties
of the energy spectrum are no more given by the impurity
configuration, but by the specular scattering at the boundaries of the
sample\cite{levy91,ullmo93,noat98}.  The orbital
magnetism of these systems should be controlled by the regular or
chaotic nature of the billiard, and must be understood in the light of
the quantum chaos theory.  Another point is the connection between the
persistent currents and the zero temperature decoherence: it has been
proposed that the anomalously high amplitude of the average current
may be related to the decoherence of the electron due to rf
environment\cite{kravtsov93,kravtsov00}.  It should also be
interesting to study thermodynamic properties different from the
persistent current.  One  example is the specific heat of
mesoscopic samples.  In a equivalent way to persistent currents,
the specific heat should oscillate with the magnetic flux.  However, such
a measurement is certainly very difficult, as the energy involved in such a
phenomenon is again on the order of the Thouless energy.  Such an
experiment would imply strong improvements in the sensitivity of present
detectors.  Finally, one subject of major interest at present is the possibility
observing the Kondo effect in artificial
nanostructures\cite{goldhaber98}.  Coupling this with a persistent
current measurement should allow to probe directly the reality and the
extension of the Kondo cloud\cite{ferrari99,kang00,affleck01}.

\end{document}